%% file: manuscript.tex
\documentclass[english,11pt]{article}

\usepackage{fullpage} 
\usepackage[section]{placeins} 
\usepackage[onehalfspacing]{setspace} 

\usepackage{amsmath} 
\usepackage{lmodern}
\usepackage[T1]{fontenc}
\usepackage[latin9]{inputenc}
\usepackage{textcomp} 
\usepackage{gensymb} 

\usepackage{booktabs} 
\usepackage{caption} 
\usepackage{subcaption} 
\usepackage{pdflscape} 
\usepackage[flushleft]{threeparttable} 

\usepackage{bbm}

\usepackage{graphicx}
\usepackage[dvipsnames]{xcolor} 

\usepackage[sort&compress,comma,authoryear]{natbib}
\usepackage{placeins}

\usepackage{soul} 
\usepackage{todonotes} 

\usepackage{tikz}
    \usetikzlibrary{arrows,positioning,decorations.pathreplacing}

\usepackage{appendix}
\usepackage{chngcntr} 

\begin{document}


\title{Climate Change and Agriculture: \\ Subsistence Farmers' Response to Extreme Heat\thanks{
\protect\linespread{1}\protect\selectfont We are grateful to Robin Burgess, Andrea Guariso, Maxim Lamare, Anirban Mitra, Matt Neidell, Debraj Ray, Finn Tarp, Marcos Vera and participants in seminars at the RES Annual Conference, the CEPR Workshop in Development (TCD/LSE), Grantham Institute (LSE), G\"{o}ttingen University - GlaD, Birkbeck University, IDB, IMF, RHUL, NWDW, NEUDC, NCDE, European Development Network, Workshop on Poverty, Inequality and Mobility at QMUL, Navarra Center for International Development, Kent Development Economics workshop, Paris Sorbonne, ZEF-Bonn, EPIC, University of Wisconsin-Madison, University of Manchester, Academia Nacional de Ciencias Econ\'{o}micas Argentina, and Universidad del Pac\'{i}fico for useful comments and suggestions. }
}

\author{Fernando M. Arag\'{o}n\thanks{
\protect\linespread{1}\protect\selectfont Department of Economics,
Simon Fraser University; email: faragons@sfu.ca}
\and Francisco Oteiza\thanks{\protect\linespread{1}\protect\selectfont Department of Social Science, UCL Institute of Education, 
email: f.oteiza@ucl.ac.uk}
\and Juan Pablo Rud\thanks{\protect\linespread{1}\protect\selectfont Department of Economics,
Royal Holloway, University of London and Institute of Fiscal Studies; email: juan.rud@rhul.ac.uk}
}

\date{February 2019}

\maketitle

\begin{abstract}
This paper examines how subsistence farmers respond to extreme heat. Using micro-data from Peruvian households, we find that high temperatures reduce agricultural productivity, increase area planted, and change crop mix. These findings are consistent with farmers using input adjustments as a short-term mechanism to attenuate the effect of extreme heat on output. This response seems to complement other coping strategies, such as selling livestock, but exacerbates the drop in yields, a standard measure of agricultural productivity. Using our estimates, we show that accounting for land adjustments is important to quantify damages associated with climate change.

 
 
\end{abstract}

\vspace{0.2in}


\newpage

 \section{Introduction}

A growing body of evidence suggests that extreme temperatures have negative effects on crop yields.\footnote{See for instance, \citet{schlenker2005will,schlenker2006impact,Deschenes2007,Lobelletal,burke2015global,carleton2016social,chen2016impacts,zhang2017economic}.} Based on these findings, current estimates suggest that climate change will bring dramatic shifts in agriculture: a global warming of 2\celsius, as in the most optimistic forecasts, would reduce agricultural output by almost 25\% \citep{ipccsynth}. Among those exposed to this shock, the rural poor in developing countries are probably most vulnerable. They are located in tropical areas, where the changes in climate will occur faster and be more intense, and their livelihoods are more dependent on agriculture.

Given these potentially disruptive effects, it is extremely important to understand possible margins of adjustment and the scope for mitigation. Some studies suggest that a possible response to climate change would be the re-allocation of economic activity, in the form of migration, changes in trade patterns or sectoral employment \citep{feng2012climate,costinot2012evolving,colmer2016weather}. 
Other studies, based on farmers' self-stated adaptive strategies, emphasize changes in consumption and savings as potential temporary responses \citep{gbetibouo2010modelling,DiFalco01042011,hisali2011adaptation}. Less is known about the potential for productive responses (i.e., changes in input use and agricultural practices), to attenuate the adverse effects of extreme temperatures. Existing studies from the U.S. and India find that extreme heat does not seem to affect crop mix or input use and that yields are affected by long-term changes in climate patterns \citep{BurkeEmerick,guiteras2009impact}. These findings have been interpreted as evidence that farmers do not engage in long-run productive adaptation. 
   
This paper examines how subsistence farmers behave when exposed to extreme temperatures. Our main contribution is to show that short-run productive responses are important, as farmers increase the intensity of input use, in particular land.
To the best of our knowledge, this margin of adjustment to extreme temperatures has not been documented before. It has, however, significant implications for the quantification of predicted economic losses due to climate change, and for understanding the potential long-term effects of weather shocks.

Our empirical analysis combines survey microdata from Peruvian farming households with weather data from satellite imagery. We examine the relationship between temperature and agricultural outcomes such as total factor productivity (TFP), yields, output, and input use. Similar to recent studies of the effect of temperature on agriculture, we use a panel data approach exploiting within-locality variation in weather.
Our approach has, however, at least three advantages over existing studies. First, the granularity of our data allows us to observe effects and responses to extreme heat at the farm level, instead of aggregated at district or county level.\footnote{A similar approach has been used for manufacturing plants in China in \citet{Zhang2017}.} This feature allows us to study a richer set of agricultural outcomes, heterogeneous responses, and other coping strategies, like the sale of disposable assets. 
Second, we study a population (i.e., subsistence farmers) that has been neglected in existing studies, but which comprises a large fraction of the rural poor around the world. 
Finally, our approach uses publicly available household and satellite data and thus can be replicated in other contexts lacking ground-level weather stations.

We find that extreme heat \textit{increases} area planted. The magnitude is economically significant: one standard deviation increase in our measure of extreme heat is associated with a 6\% increase in land used. Consistent with the additional land being planted with a different crop mix, we find that extreme heat increases the quantity harvested (in absolute and relative terms) of tubers (i.e., potatoes). We also find suggestive evidence of increments in the use of domestic, including child, labor on the farm. This increase in input use occurs despite high temperatures reducing agricultural productivity, and partially offsets the drop in total output.

We interpret these findings as evidence that subsistence farmers respond to extreme temperatures by increasing input use. This productive adjustment attenuates undesirable drops in output and consumption. This interpretation is consistent with agricultural household models with incomplete markets  \citep{de1991peasant,taylor2003agricultural}. In these models, production and consumption decisions are not separable. Thus, farmers may resort to more intensive use of non-traded inputs, like land and domestic labor, to offset the impact of negative income or productivity shocks. This margin of adjustment may be particularly relevant for farmers in less developed countries due to the presence of several market imperfections and limited coping mechanisms.

We also examine several ex-post coping mechanisms previously identified in the literature on consumption smoothing, such as migration, off-farm labor, and disposal of livestock \citep{BEEGLE,BANDARA,Kochar,Munshi,Rosenzweig_Stark,Rosenzweig_Wolpin}. Consistent with previous studies, we find that households reduce their holdings of livestock after a negative weather shock and seem to increase hours working off the farm. Interestingly, the increase in land use as a response to extreme heat occurs even among farmers that resort to consumption smoothing strategies. This finding suggests that productive responses to extreme temperatures remain important to traditional farms even if they have alternative risk-coping instruments at hand.

Our findings have two important implications. First, they suggest a potential dynamic link between weather shocks and long-run outcomes. If the increase in land use comes at the expense of investments (such as fallowing), then this short-term response could affect future land productivity. A similar argument could be made about child labor. While we are unable to examine these implications due to data limitations, future research should explore these links more closely. Second, this farmer response may affect estimations of the damages of climate change on agricultural output. These estimates are usually based on the effect of temperature on crop yields \citep{Deschenes2007}. This is a correct approach if land use is fixed. In that case, changes in crop yields are the same as changes in output. However, if area planted increases with temperature, then using crop yields would overestimate the resulting loss in output. 

To illustrate this point, we use our results to predict damages of climate change by the end of the century under two standard scenarios (RCP45 and RCP85). Using the effect of temperature on yields, as in the existing literature, suggests output losses of 5-9\% under different scenarios. In contrast, taking into account changes in land use, we obtain much smaller losses of 0.6 to 1.2\%.

The rest of this paper is organized as follows. Section \ref{section_background} describes the context and the analytic framework. Section \ref{section_methods} discusses the data and the empirical strategy. Section \ref{section_results} presents our main results and robustness checks. Section \ref{section_other_coping}  examines other other coping mechanisms, while Section \ref{section_CC} discusses the implication of our findings for estimating climate change damages. Section \ref{section_conclusion} concludes.


\section{Background}\label{section_background}

\subsection{Subsistence farming in Peru}\label{section_subsistence}
Our empirical analysis focuses on subsistence farmers from rural Peru. In 2017, the last year of our study, 24\% of the working population was employed in agriculture, but the sector only accounted for 7\% of the GDP \citep{INEI2018}.  It is, in other words, a sector of very low productivity, with many characteristics in common with subsistence farming in other developing countries: it is mainly composed by small productive units, with low capital intensity, and low levels of technology adoption \citep{velazco2012caracteristicas}.

Table \ref{table_summary} presents some key summary statistics of the farmers in our sample and defines the setting for our analysis.\footnote{Data sources and variable definitions are described in Section \ref{section_data}} Most farmers are poor and depend on agriculture as their main source of livelihood. The incidence of poverty in our sample of farmers is around 50\%. For comparison purposes, a similar methodology shows that poverty over the whole of Peru during the period of analysis was 21.6\%. The average farm is around 2 hectares, has a low degree of specialization and uses practices akin to traditional, rather than industrial, farming. They rely on domestic labor (including child labor), cultivate a variety of crops instead of monocropping, and leave some land uncultivated. While this feature is consistent with fallowing and crop rotation, a large part of the land is not used for productive purposes (planting or fallowing).

\begin{figure}[!h]
	\centering
	\caption{Hectares planted by calendar month and climatic region} \label{fig_calendar}
	\includegraphics[width=15cm]{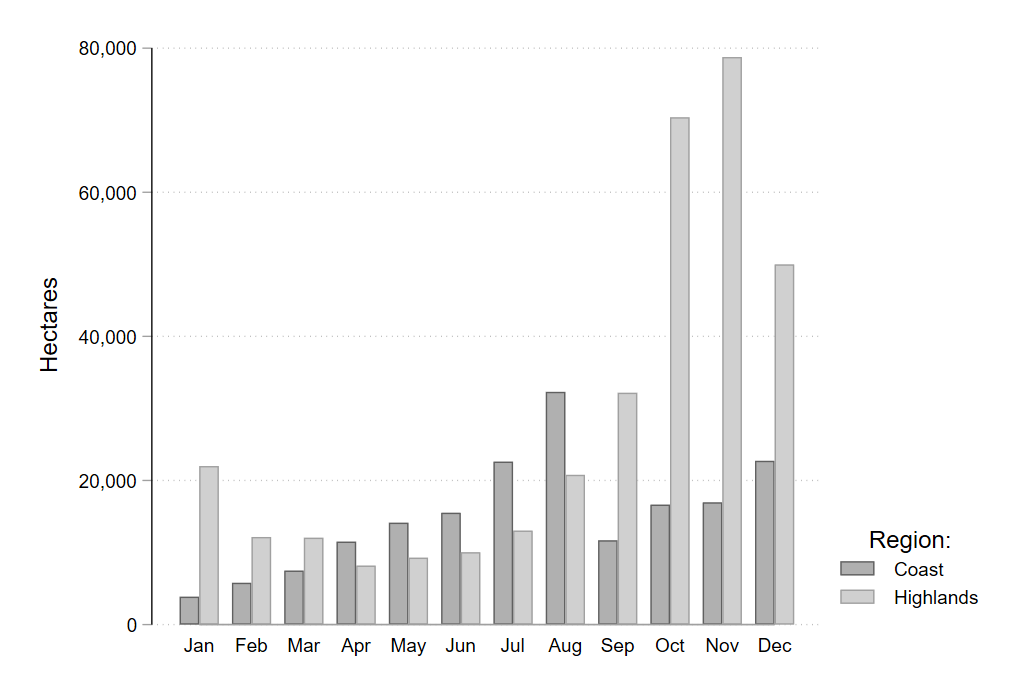}
	\parbox{14cm}{\footnotesize{\textit{Notes}: Total hectares planted by calendar month, for the crops harvested during the 2017 growing season. Planting period goes from January 2015 to May 2017. 
	\emph{Source:} National Agricultural Survey of Peru, 2017.}}
\end{figure}

Figure \ref{fig_calendar} shows the number of hectares planted by calendar month, during 2016-2017. As one can see, most planting occurs during October-March. These months correspond to spring and summer in the Southern Hemisphere and are considered the main growing season in Peru. However, planting is not a one-off activity as it persists throughout the year. This feature suggests that farmers have some margin to adjust their input use during the agricultural year.  
  
Our study concentrates on two climatic regions: the coast and the highlands.\footnote{Peru has three main climatic regions: the coast to the west, the Andean highlands, and the Amazon jungle to the east.} These two regions exhibit different climate driven by their proximity to the sea and altitude. The coast is a narrow strip extending from the seashore up to 500 meters above sea level (masl). It has a semi-arid climate, with warm temperatures and little precipitation.  The highlands extends from 500  up to almost 7,000 masl, albeit most agriculture stops below 4000 masl. It has a much cooler and wetter climate, with seasonal precipitations in spring and early summer. 

These climatic differences are associated with different agricultural practices: coastal farmers are more reliant on irrigation, while agriculture in the highlands is mostly rainfed. Coastal farmers are also less likely to be poor and have a different crop mix, cultivating a larger share of fruits and cereals. While these regional differences do not affect the key results in our analysis (see Section \ref{section_robustness}), they have important implications in terms of the potential effects of greater temperatures due to climate change. 

\begin{table}[!h]
	\centering
	\begin{threeparttable}
		\caption{Summary statistics (ENAHO 2007-2015)}
		\label{table_summary}  
		 \input{content/table_summary.tex}
		\begin{tablenotes}
			\footnotesize
			\item \emph{Notes:} Output and livestock value measured in 2007 USD. Land is measured in hectares. Temperature is measures in Celsius degrees. HoH= household head. HH= household. DD=degree days. HDD = harmful degree days.
		\end{tablenotes}
	\end{threeparttable}
\end{table}

\FloatBarrier

 \subsection{Analytical framework}\label{section_model}
 
 This section develops a simple framework to examine how subsistence farmers adjust their production decisions as a response to extreme heat. To this end, we follow standard agricultural household models in the development literature \citep{de1991peasant,Benjamin,taylor2003agricultural} where households make simultaneous, potentially interrelated, consumption and production decisions during the growing season. 

Consider a producer-consumer household. 
The household has endowments of land and labor, $L^{e}$ and $T^{e}$. These endowments can be used for production or ``consumed'' in non-productive activities (e.g., leisure).
Household's utility is $U(c,l,t)$, where $c$ is consumption of a market good, while $l$ and $t$ are the amounts of labor and land used in non-productive activities. 
Households obtain income by selling their endowments and by producing an agricultural good. Production is defined by function $F(A,L,T)$, where $T$ and $L$ are the amount of labor and land used, and $A$ is farmer's total factor productivity.

$A$ is a productivity shifter that captures the idea that farmers using identical inputs can have different levels of output due, for instance, to different farming skills, soil quality, or exposure to weather shocks.\footnote{In our context, we assume that capital such as irrigation, if used at all, is fixed.}
Consistent with previous studies on the relation between crop yields and temperature, we assume that extreme heat has a detrimental effect on productivity.\footnote{See for example \citet{Schlenker15092009}, \citet{BurkeEmerick}, \citet{auffhammer2012climate}, \citet{hsiang2010temperatures}, \cite{hsiang2016climate}, among others.}      

Each growing season, the household maximizes utility by choosing the amount of land and labor allocated to productive and non-productive uses.
We consider that both land and labor are variable inputs. This assumption is driven by the  observation that, among subsistence farmers, planting is not a one-off activity, but instead it is spread throughout the year (see Figure \ref{fig_calendar}).\footnote{Note that multi-cropping practices, combined with the availability of uncultivated land, implies that both inputs and outputs are flexible throughout the season, during which $A$ is realized.}
Finally, we assume that both the utility and the production functions are increasing and strictly concave.

\subsubsection{Household responses to negative productivity shocks}

 If input markets exist and are well functioning, we can study consumption and production decisions separately \citep{Benjamin}. This separation result is driven by the possibility to trade. Thus, the household's demand and supply of inputs for production and consumption need not be identical to its endowments.
 The farmer's use of inputs on the farm can then be analyzed by solving the profit maximization problem  $\underset{T,L}{\text{max}} \quad \pi = p f(A,L, T) - wL - rT$, were $p$, $w$ and $r$ refers to output and input prices. 
The standard solution is the unconditional demand for inputs $L^{*}(A,p,w)$ and $T^{*}(A,p,w)$. In this context, a farmer's response to negative productivity shock, such as extreme heat, is unequivocal: she will \textit{reduce} the amount of land and labor used in her farm. 

This prediction can change in the case of incomplete markets. 
To illustrate this, consider a case in which the only input is land, and there are no input markets. In this simplified setting, the farmer's problem becomes:
\begin{equation*}\label{eq_problem}
    \begin{aligned}
    & \underset{T}{\text{max}}
    &   &  U(c, t ) \\
    & \text{s.t.} 
    &   &  c = p F(A,T) \\
    & &   &   T +t = T^e.
    \end{aligned}
\end{equation*}

Solving this problem produces an unconditional demand for land that depends not only on prices and productivity, but also on land endowment, $T(A,p,T^e)$. Moreover, if utility is sufficiently concave (for instance if consumption levels are quite low or farmer has high risk aversion), then $\frac{d T}{d A}$ can be negative.\footnote{Taking  total derivatives to first order condition $U_c F_T = U_t$, we obtain that:
\begin{equation*}
    \frac{dT}{dA} ( F_T^2 U_{cc} + U_c F_{TT} + U_{tt}) + F_T F_A U_{cc} + U_c F_{TA} = 0.
\end{equation*}
This expression implies that $-\frac{U_{cc}}{U_c} > \frac{F_{TA}}{F_T F_A}$ is a sufficient condition for inputs to increase with a negative productivity shock, i.e. $\frac{dT}{dA}<0$. For example, for a Cobb-Douglas technology $f=AT^{\alpha}$, this condition simplifies to:
$-\frac{U_{cc}}{U_c} >1$. } 

This result suggests that, in context with imperfect input markets, negative weather shocks could result in an \textit{increase} in input use.
This occurs because the farmer uses more inputs to attenuate the fall in agricultural output, and reduce the drop in consumption. This response is akin to coping mechanisms to smooth consumption, such as selling disposable assets. The key distinction is that it involves adjustments in productive decisions. 
This prediction is relevant because subsistence farmers in rural Peru (and other parts of the developing world) likely face severe imperfections in input markets \citep{gollin2013agricultural,restuccia2008agriculture}. 






With this framework in mind, our empirical analysis focuses on examining the effect of extreme heat on input use, as well as on agricultural productivity. There are, however, other possible responses. For instance, recent work on climate change and adaptation has stressed changes in crop mix as a possible response \citep{BurkeEmerick,costinot2012evolving}. Similarly, an influential literature highlights how households can smooth consumption by migrating, increasing off-farm work, or selling cattle, among other strategies (see for instance \citet{Rosenzweig_Wolpin} or \citet{Kochar}). In the empirical section, we also examine these additional potential responses.

\section{Methods}\label{section_methods}
        \subsection{Data} \label{section_data}
 
We combine household surveys with satellite imagery to construct a comprehensive dataset containing agricultural, socio-economic, and weather variables. The unit of observation is the household-year. We restrict the sample to households with agricultural activities located in the coast and highlands.\footnote{We do not include observations from the jungle due to small sample size and poor quality of satellite data. We also drop 282 farmers from the coast and highlands reporting land holdings larger than 100 hectares.} Our final dataset consists of around 53,000 observations and spans over the years 2007 to 2015. Table \ref{table_summary} presents some summary statistics for our sample.
  
\subsubsection{Agricultural and socio-economic data}

Our main data source is repeated cross-sections of the Peruvian Living Standards Survey (ENAHO), an annual household survey collected by the National Statistics Office (INEI). This survey is collected in a continuous, rolling, basis. This feature guarantees that the sample is evenly distributed over the course of the calendar year. 
  
The survey asks the farmer to report the quantity of crops harvested in the last 12 months, as well as the size and use of parcels planted in that period. We use this information to construct measures of agricultural output and input use. To measure real agricultural output, we construct a Laspeyres index using quantity produced of each crop and baseline local prices.\footnote{As weights, we use the median price of each crop in a given department in 2007.} We calculate land used by adding the size of parcels dedicated to seasonal and permanent crops. We distinguish between domestic and hired labor. We measure hired labor using self-reported wage bill paid to external workers in the last 12 months. To measure domestic labor, we use information on household members' employment. In particular, we calculate the number of household members working in agriculture and build an indicator of child labor.\footnote{Child labor is defined as an indicator equal to one if a child living in the household aged 6-14 reports doing any activity to obtain some income. This includes helping in the family farm, selling services or goods, or helping relatives, but excludes household chores.}

This dataset has three relevant limitations. First, we do not observe the time of planting, only the total land used in the last 12 months. Second, we do not observe which specific crops are cultivated in each parcel.\footnote{We only observe total area planted and, separately, total harvests of each crop.} Since most farmers grow several crops and practice inter-cropping, we cannot calculate crop-specific yields. Finally, the information on household employment is available only for the two weeks before the interview. Given that interviews can occur all year round and labor use is seasonal, our measures of domestic labor may not reflect actual input use during the whole year. While this measurement error does not affect estimates of the effect of temperature on land use, it can affect estimates of its impact on labor use. In those cases, we address this concern by restricting the sample to farmers interviewed during the main growing season only.

The survey also provides information on socio-demographic characteristics, agricultural practices and farm conditions (such as intercropping, access to irrigation, and use of fertilizers), and geographical coordinates of each primary sampling unit or survey block.\footnote{There are around 3,800 unique coordinate points in our sample. Figure \ref{fig_mapPeru} in the Appendix depicts the location of clusters used in this study.} In rural areas, this corresponds to a village or cluster of dwellings.  We use this geographical information to link the household data to satellite imagery. We complement the household survey with data on soil quality from the Harmonized World Soil Database \citep{fao2008}. This dataset provides information on several soil characteristics relevant for crop production on a 9 km square grid.\footnote{The soil qualities include nutrient availability and retention, rooting conditions, oxygen availability, excess salts, toxicity, and workability.}
 
\subsubsection{Temperature and precipitation}
We use satellite imagery to obtain high-resolution measures of local temperature. We prefer to use satellite imagery instead of ground-level measures or gridded products, such as re-analysis datasets, due to the small number of monitoring stations (around 14 in the whole country).\footnote{Note that reanalysis datasets use ground-level readings as a main input and thus can be less precise in contexts with a low number of monitoring stations \citep{auffhammer2013global}.} We use the MOD11C1 product provided by NASA. This product is constructed using readings taken by the MODIS tool aboard the Terra satellite. These readings are processed to obtain daily measures of daytime temperature on a grid of $0.05 \times 0.05$ degrees, equivalent to 5.6 km squares at the Equator, and is already cleaned of low quality readings and processed for consistency.\footnote{MODIS validation studies comparing remotely sensed land surface temperature estimates and ground, \emph{in situ}, air temperature readings found discrepancies within the 0.1-0.4 \celsius\ range \citep{Coll2005,Wan2008,Coll2009}.}

The satellite data provides estimates of land surface temperature (LST) not of surface air temperature, which is the variable measured by monitoring stations. For that reason, the reader should be careful when comparing the results of this paper to other studies using re-analysis data or station readings. LST is usually higher than air temperature, and this difference tends to increase with the roughness of the terrain. However, both indicators are highly correlated \citep{mutiibwa2015land}. 

We complement the data on temperature with information on local precipitation. We use data from the Climate Hazards Group InfraRed Precipitation with Station data (CHIRPS) product \citep{chirps}. CHIRPS is a re-analysis gridded dataset that combines satellite imagery with monitoring station data. It provides estimates of monthly precipitation with a resolution of $0.05 \times 0.05$ degrees.

\begin{figure}[h]
   	\centering
   	\caption{Distribution of daily average temperature by growing season} \label{fig_distribution}
   	\includegraphics[width=15cm]{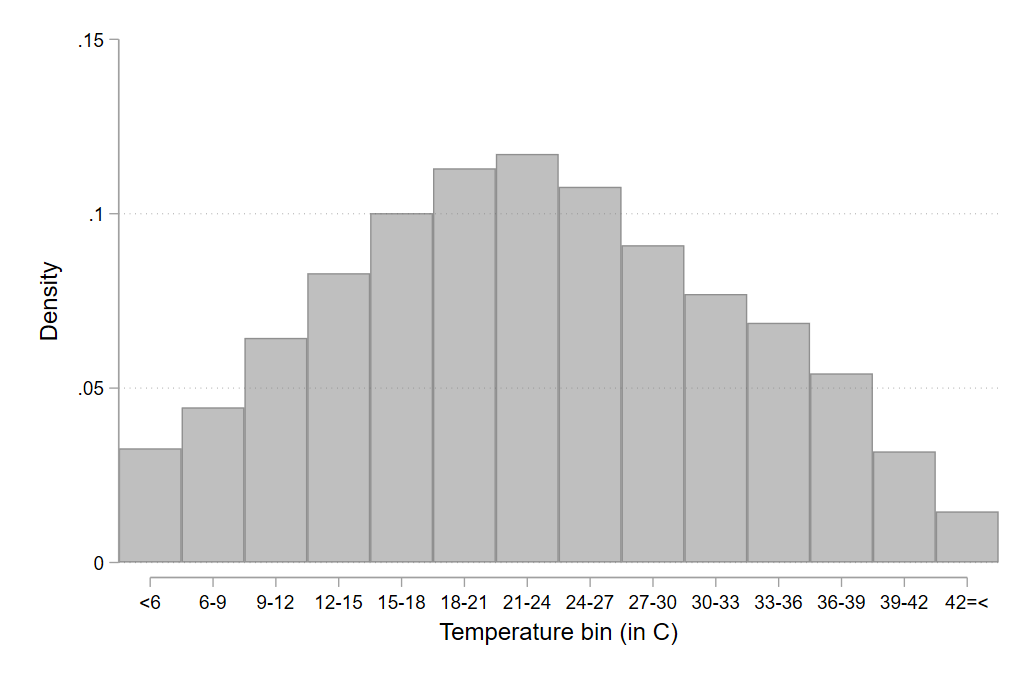}
   	\footnotesize
	\vspace{10pt}
   	\parbox{15cm}{\footnotesize{Notes: density of daily temperatures during the last completed growing season (i.e., October to March). The unit of observation is farmer-growing season.}}
\end{figure}

To link the weather and household data, we attribute to a given household the weather conditions in the cell overlapping its coordinates. Then, we aggregate weather data (which have daily and monthly frequency) to obtain measures of exposure to weather during a given agricultural year. In our baseline specification, we focus on exposure to weather during the last completed growing season. The growing season is the period in which most of planting and crop growth occurs. As shown in Section \ref{section_subsistence}, even though planting is a year-round activity, it is particularly concentrated in spring and summer. We use this period as our definition of growing season.\footnote{We define the growing season as months October to March. In Section \ref{section_robustness}, we check the robustness of our results to alternative ways to aggregate weather over time, such as by climatic season or during last 12 months.} Figure \ref{fig_distribution} shows the distribution of temperatures observed during the last completed growing season for our whole sample.\footnote{Figure \ref{fig_distbygs} in the Appendix, shows the average distribution of temperatures by growing season, and shows that the distribution is mostly stable over the time of our study.}

\subsection{Empirical strategy}

The empirical analysis aims to study how farmers respond to extreme heat. Based on the discussion in Section \ref{section_model}, we focus on productive adjustments, such as changes in input use. To study this response, we estimate reduced-form unconditional factor demands linking input use to weather shocks.

In a standard production model, unconditional factor demands are a function of total factor productivity (TFP), and agricultural prices. In the presence of imperfect input markets, they can also be affected by household endowments.\footnote{For instance, in the extreme case of no input markets, input use would be proportional to input endowments.} In this context, weather conditions, such as temperature and precipitation, enter into the factor demand through their effects on $A$.

We approximate the reduced-form factor demand using the following log-linear regression model:
     \begin{equation}\label{estimated_model}
     \ln y_{ijt} =  g(\gamma, \omega_{it}) +  \phi Z_{i} + \rho_{j}+ \psi_t +  \epsilon_{ijt},
     \end{equation}
where the unit of observation is farmer $i$ in district $j$ and growing season $t$. $y$ is our measure of input use and $g(\gamma, \omega_{it})$ is a non-linear function of temperature and precipitation ($\omega_{it}$). The parameter of interest is $\gamma$: the reduced-form estimates of the effect of weather shocks on input use. Note that our specification exploits within-district variation. Thus we cannot estimate the effect of climate, but only of weather shocks. This approach is similar to the panel regressions used in recent studies of the effect of climate on economic outcomes \citep{Dell2014}.

$Z_{i}$ is a vector of farmer characteristics, $\rho_{j}$ is a set of district fixed effects, and $\psi_t$ are climatic region-by-growing season fixed effects.\footnote{A district is the smallest administrative jurisdiction in Peru and approximately half the size of the average U.S. county. Our sample includes 1,320 districts out of a total of 1,854.}
These control variables proxy for both determinants of TFP as well as other drivers of input use.
$Z_{i}$ includes possible drivers of TFP such as indicators of soil quality, household head's education, age, and gender, as well as measures of input endowments like land owned and household size,
$\psi_t$ controls for common productivity shocks but, to the extent that agricultural markets are national, also for agricultural prices. Similarly,  $\rho_{j}$ accounts for location-specific determinants of productivity, such as climate and soil quality, but can also control for other time-invariant determinants of input use, like proximity to markets. 
 
We modify our baseline specification \eqref{estimated_model} to examine the relation between temperature and agricultural productivity. We use two approaches. First, we follow the existing literature and use yields (i.e., output per unit of land) as our measure of (land) productivity. 
A limitation of this approach is that yields are a measure of partial productivity that reflect changes in TFP and land used. This is not an issue when land is fixed, but can overestimate the effect of extreme heat on productivity if farmers adjust land. 

As a second approach, we estimate a production function. Assuming a Cobb-Douglas specification we modify our baseline specification by using log of output as outcome and controlling for log of input use.\footnote{Assuming a Cobb-Douglas production function $Y_{ijt}=A_{ijt} T_{it}^{\alpha} L_{it}^{\beta}$, applying logarithms, and defining $A=\exp(g(\gamma, \omega_{it}) +  \phi Z_{i} + \rho_{j}+ \psi_t + \epsilon_{ijt})$  we obtain the following regression model:

\begin{equation*} 
    \ln Y_{ijt}=\alpha \ln T_{it} + \beta \ln L_{it} + g(\gamma, \omega_{it}) +  \phi Z_{i} + \rho_{j}+ \psi_t + \epsilon_{ijt},
    \end{equation*}   where $Y$ is agricultural output, and $T$ and $L$ are quantities of land and labor. } This approach allow us to estimate directly the effect of extreme heat on TFP. However, it comes at the cost of imposing parametric assumptions and, potentially, creating an endogeneity problem due to omitted variables affecting both TFP and input use. Consistently with the analytical framework proposed in Section \ref{section_model},  we address this issue by using endowments, such as household size and owned land, as predictors for input use in an instrumental variable approach.

\subsubsection{Modeling the relation between weather and productivity $g(\gamma, \omega_{it})$} 

Similar to previous work, we model the relation between weather and agricultural productivity as a function of cumulative exposure to heat and water.\footnote{See for instance \citet{schlenker2006b} and \citet{schlenker2006impact}.}
In particular, we construct two measures of cumulative exposure to heat during the growing season (i.e., spring and summer): average degree days (DD) and harmful degree days (HDD). 

DD measures the cumulative exposure to temperatures between a lower bound, usually 8\celsius\, up to an upper threshold  $\tau$, while HDD captures exposure to temperatures above $\tau$. The inclusion of HDD allows for potentially different, non-linear, effects of extreme heat. Formally, we define the average DD and HDD during the growing season as:
\begin{eqnarray*}
DD = \frac{1}{n} \sum_{d=1}^n (h_d - 8) \mathbbm{1}{(8 \leq h_d \leq \tau)  }  \\
HDD = \frac{1}{n} \sum_{d=1}^n (h_d - \tau_{high})) \mathbbm{1}{( h_d > \tau)  }, 
\end{eqnarray*}

where $h_d$ is the average daytime temperature in day $d$ and $n$ is the total number of days in a growing season with valid temperature data. Note that we do not calculate \textit{total} degree days, but instead the \textit{average} degree days. This re-scaling makes interpretation easier and help us address the issue of missing observations due to satellite swath errors.
 
\begin{figure}[h]
    \centering
    \caption{Non-linear relationship between temperature and agricultural yields} \label{fig_bin_peru}
    \includegraphics[width=15cm]{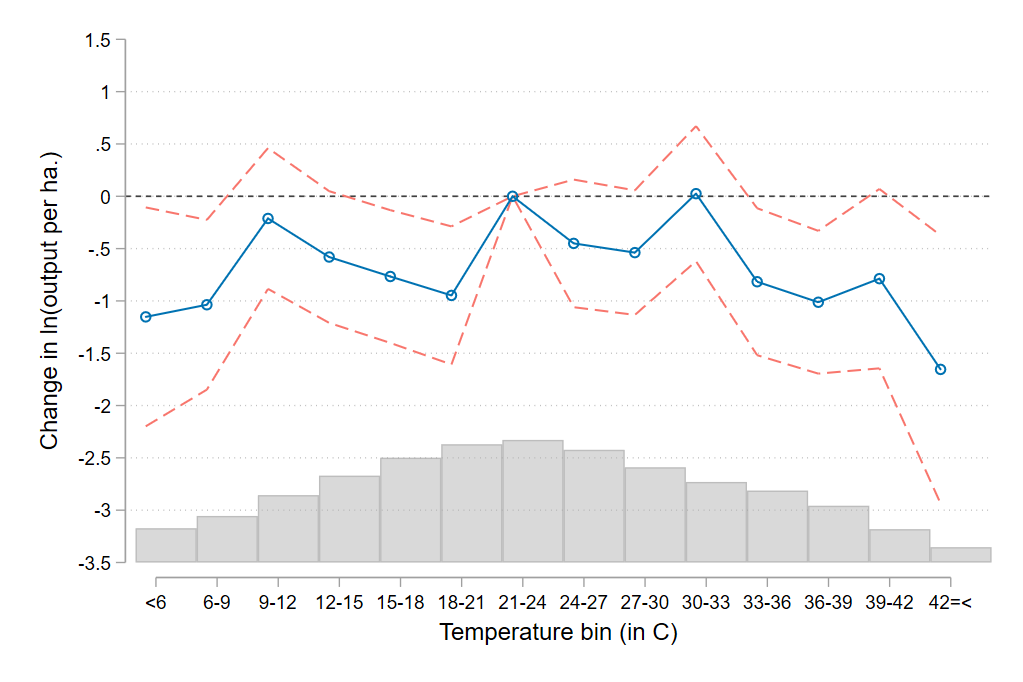}
    \vspace{10pt}
    \parbox{15cm}{\footnotesize{\emph{Notes:} Points represent coefficient estimates of the effect of increasing the share of days in the growing season in each of the temperature bins, relative to the 27-30\celsius\ bin, on log of output per ha. Bin labels denote the lower (included) and upper (excluded) bounds. For example, bin 27-30 measures the fraction of the growing season spent on days with temperatures above or equal to 27\celsius\ and lower than 29\celsius. Dashed lines show the 95\% confidence interval.}}
\end{figure}

A key issue is to define the value of $\tau$. Previous studies in U.S. set this value between 29-32\celsius\ \citep{schlenker2006b,Deschenes2007}. These estimates, however, are likely to be crop  and context dependent and hence might not be transferable to our case.\footnote{In addition to differences in crop mix and agricultural technology, we use a different measure of temperature (i.e., land surface temperature). These factors make previous estimates not applicable to our case study.} For that reason, we prefer to use a data-driven approach. To do so, we estimate a flexible version of equation \eqref{estimated_model} using log of output per hectare as outcome variable and replacing $g(.)$ with a vector of variables measuring the proportion of days in a growing season on which the temperature fell in a given temperature bin.\footnote{This specification is similar to the one used by \citet{burgess2017weather} to study the effect of weather on mortality. Based on the distribution of temperatures in the Peruvian case, we define 14 bins: $<6$\celsius, $\geq 42$\celsius, and twelve 3\celsius-wide bins in between. Our omitted category is the temperature bin 21-24 \celsius.} 
The results, displayed in Figure \ref{fig_bin_peru} suggest that point estimates become negative for temperatures above 33 \celsius. We use this temperature as our preferred $\tau$ in our baseline specification.\footnote{As a robustness check, we also estimate $\tau$ using an iterative regression method similar to those used by \cite{schlenker2006impact}. We ran 17 regressions with different DD/HDD thresholds ranging from 26\celsius\ to 42\celsius\ and compared their model fit. The results, in Figure \ref{fig:threshold_test} suggests optimal temperatures in the slightly lower 30-32 \celsius\ range. To ensure that our choice of $\tau$ does not drive our main results, in Figures \ref{fig:fig_robustness_threshold1} and \ref{fig:fig_robustness_threshold2} in the Appendix we plot the point estimates of the HDD coefficients for the range of $\tau$ mentioned above. Reassuringly, point estimates are of similar size, magnitude, and precision between the 26-35\celsius\ interval.}

We measure exposure to precipitation using the average daily precipitation (PP) during the growing season and its square. 
With these definitions in mind, we parametrize the function relating weather to productivity $g(\gamma, \omega_{it})$ as:
\begin{equation*}
    g(\gamma, \omega_{it}) = \gamma_0 DD_{it} + \gamma_1 HDD_{it} + \gamma_2 PP_{it} + \gamma_3 PP^2_{it}.
\end{equation*}

\section{Main results}\label{section_results}

This section presents our main empirical results on farmers' responses to extreme heat. We begin by documenting the negative effect of high temperatures on agricultural productivity. Then we examine possible productive responses, such as changes in input use. Finally, we study the interaction of these productive responses with other coping strategies already discussed in the consumption smoothing literature.

\subsection{Temperature and agricultural productivity}

We start by examining whether extreme heat is indeed a negative productivity shock. Previous studies in other countries suggest a non-linear relation between temperature and agricultural productivity: at high temperatures additional heat is detrimental to crop yields.\footnote{See, for instance, 
 \citet{auffhammer2012climate}, \citet{guiteras2009impact}, \citet{burgess2017weather}, \cite{burke2015global}, \cite{BurkeEmerick}, \citet{Schlenker15092009}, \citet{Lobelletal}.}
Figure \ref{fig_bin_peru} corroborates this result in the Peruvian context using the temperature bin approach.

Table \ref{table_main} presents our results using our baseline specification that measures exposure to heat using degree days (DD) and harmful degree days (HDD) averaged over the main growing season (i.e., spring and summer). Column 1 uses agricultural yields ($Y/T$) as a proxy for productivity. A limitation of this approach is that using yields confounds impacts on both TFP and land use. For that reason, in columns 2 and 3 we estimate a Cobb-Douglas production function, i.e., output conditional on input use, to examine the effect of temperature of total factor productivity (TFP). We estimate the production function using both OLS and 2SLS. In the latter case, we instrument input use with household endowments (i.e., area of land owned and household size).

Our estimates suggest that extreme heat has a negative effect on agricultural productivity. The magnitude of the effect is economically significant: the most conservative estimate suggests that each additional average HDD results in a 7\% decrease in agricultural productivity. To put this figure in perspective, note that climate change scenarios discussed in Section \ref{section_CC} envisage that, by the end of this century, the average number of HDD over the growing season could increase between 0.64 and 1.32, while the already warm Coast would experience increments between 3 to 5 HDD.  
    
What happens with total output? Consistent with a negative productivity shock, we find that extreme heat reduces agricultural output (column 4). However, the magnitude of this effect is smaller than for TFP or yields, and we cannot reject the null hypothesis at standard levels of confidence. This finding is suggestive of responses (such as changes in production decisions) that attenuate the effect of the productivity shock on total output. 

 \begin{table}[h!]
      \centering
      \begin{threeparttable}
          \caption{Temperature, agricultural productivity and output}
          \label{table_main} 
          \input{content/table_main.tex}
          \begin{tablenotes}
              \footnotesize
              \item \emph{Notes:} Standard errors (in parenthesis) are clustered at the district level. Stars indicate statistical significance: \textit{*p \textless 0.10, ** p \textless 0.05, *** p \textless 0.01}. All specifications include  district, month of interview, and climatic region-by-growing season fixed effects, and farmer controls such as:  household head characteristics (age, age$^2$, gender, and level of education), indicators of soil quality from \cite{fao2008} (nutrient availability, nutrient retention, rooting conditions, oxygen availability, salinity, toxicity and workability) and the share of irrigated land. Input controls: log of area planted, number of household members working in agriculture, and amount spent on hired labor. Instruments for domestic labor and area planted: log of household size and area of land owned. First stage joint significance F-test is 466.7.   
          \end{tablenotes}
      \end{threeparttable}
  \end{table}

\subsection{Productive responses: changes in input use}

We examine changes in input use as a potential margin of adjustment to high temperatures. In our main set of results, we focus on changes in land use, both in terms of area planted and crop mix. Our focus on land stems from its importance as an agricultural input and because, in many contexts, 
it is subject to severe market imperfections, such as ill-defined property rights. Moreover, we have reasonably good measures of land use, but more limited information on other inputs, such as labor.

Table \ref{table_landuse} presents our main results. We find a positive and statistically significant effect of HDD on area planted (column 1). An increase in HHD of 1 degree is associated with an increase of almost 6\% in the total area planted. This estimate already controls for endowments, such as the total area of land available, and thus is not simply picking up changes in the size composition of farmers. The increase in land used is sizable and partially explains why, despite its documented negative effects on agricultural productivity, extreme heat has a small and insignificant effect on total output. It also explains why the estimated effect of HDD on yields (Y/T) is larger than on total factor productivity (TFP) (see Table \ref{table_main}).

Columns 2 to 4 examine the effect of extreme heat on crop mix. In our context, farmers practice multi-cropping: the average farmer grows almost six different crops.\footnote{In our sample, less than 10\% of farmers report growing only one crop. Multi-cropping is a common practice among subsistence farmers across the developing world, and is in stark contrast with the modern agricultural practices of the U.S. and other developed countries, which mostly practice mono-cropping.} To study effects on crop mix, we group crops in two categories: tubers (mostly potatoes) and other crops. Tubers are the most important crop among Peruvian subsistence farmers and account for almost 30\% of the value of agricultural output and 15\% of the area planted. 
  
We find that extreme heat increases the quantity (in absolute and relative terms) of tubers harvested. We interpret these findings as suggestive evidence that the additional land is planted with a higher share of tubers. Hence farmers adjust their use of land, both in terms of area planted and crop composition, as a response to extreme heat. These results complement recent studies that examine the role of changes in crop mix as a possible adaptation to climate change \citep{BurkeEmerick,colmer2016weather}.

There are, however, two important caveats. First, we do not observe the area planted with different crops, only the amount harvested. Thus, we are unable to disentangle the effect of extreme heat on planting decisions from different crop sensitivities to temperature. That said, we can rule out that our results are only reflecting less sensitivity of tubers to extreme heat: in that case, we would observe an increase in output share, but a reduction in absolute terms. Second, our results do not necessarily mean that tubers are more resilient to heat than other crops. Farmers could prefer tubers because they have a more flexible planting schedule, provide more calories per unit of land, have fewer requirement of other inputs (like fertilizers), or are less risky.\footnote{For instance, \citet{dercon1996} documents in the Tanzanian context, that farmers manage risk by planting less profitable, but more reliable, crops like sweet potatoes.}

  \begin{table}[h!]
        \centering
        \begin{threeparttable}
            \caption{Temperature and land use}
            \label{table_landuse} 
            \input{content/table_landuse.tex}
            \begin{tablenotes}
                \footnotesize
                \item \emph{Notes:} Standard errors clustered at the district level (in parenthesis). Stars indicate statistical significance: \textit{*p \textless 0.10, ** p \textless 0.05, *** p \textless 0.01}. All specifications include  district, month of interview, and climatic region-by-growing season fixed effects, and the same farmer controls as baseline regression in Table \ref{table_main}. Endowment controls: log of household size and area of land owned.
            \end{tablenotes}
        \end{threeparttable}
    \end{table}
    
\paragraph{Timing}

Do the effect and responses to extreme heat vary according to the time at which extreme temperatures are experienced? Answering this question is relevant to understand the observed phenomena better and predict impacts more accurately. For instance, effects could vary if crops are more sensitive to extreme heat at some stages of development (sowing, harvesting) than others, or if farmers face time-varying constraints to adjust to these shocks (i.e., due to seasonal crop or input suitability). Alternatively, we might be observing a delayed response from farmers to extreme temperatures in previous agricultural seasons, not a response to a contemporaneous shock.
 
To examine this issue, we first restrict our sample to those farmers interviewed during the fall or winter months (April to September, in the southern hemisphere). As mentioned before, although planting and harvesting are year-round activities, the most important planting period (in terms of area) corresponds to spring and summer, the growing season months. Thus, our sample restriction allows us to focus on those farmers who have already completed most of their annual land use decisions.\footnote{Recall that interviewers ask about the total land used in agriculture over the past 12 months} Then, we construct separate measures of weather for each of the last four seasons (i.e., fall, winter, spring, and summer). Specifically, if a household is interviewed during the fall or winter of year $t$, we match each observation with the weather outcomes in that location during the fall, winter and spring of year $t-1$ (April to December), and for the summer of year $t$ (January to March). This procedure effectively summarizes the weather conditions over the 12 months previous to the end of the last growing season.

\begin{figure}[h!]
    \centering
    \caption{Effect of exposure to HDD by season} \label{fig_timing}
    \begin{subfigure}[b]{0.7\textwidth}
        \includegraphics[width=\textwidth]{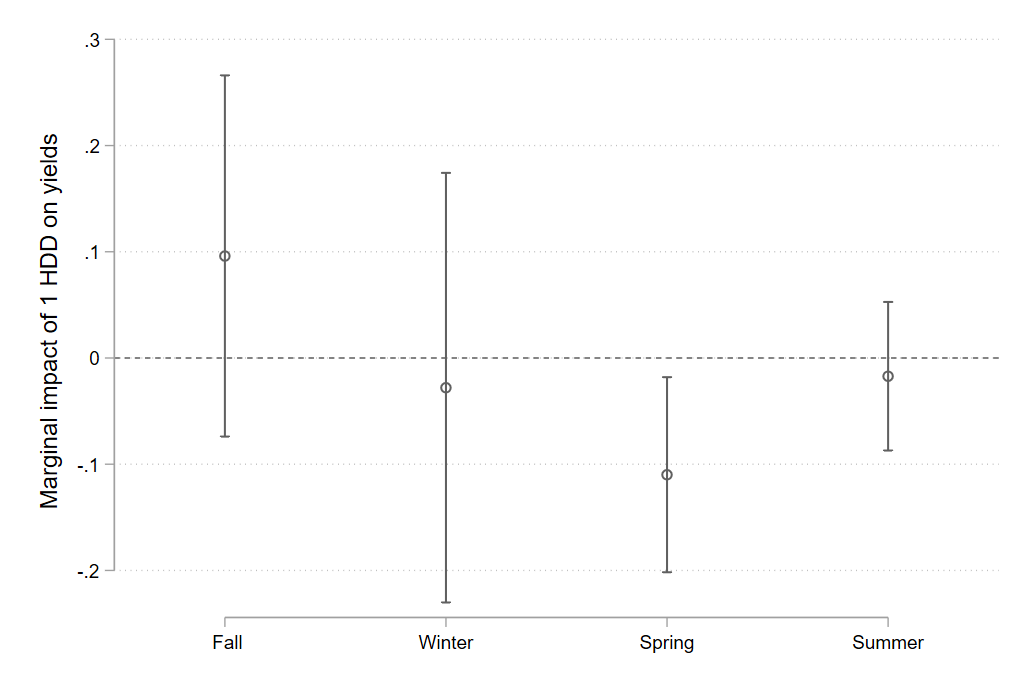}
        \caption{Impacts on ln(output per hectare planted)}
        \label{fig:timing_YT}
    \end{subfigure}
    ~ 
    \begin{subfigure}[b]{0.7\textwidth}
        \includegraphics[width=\textwidth]{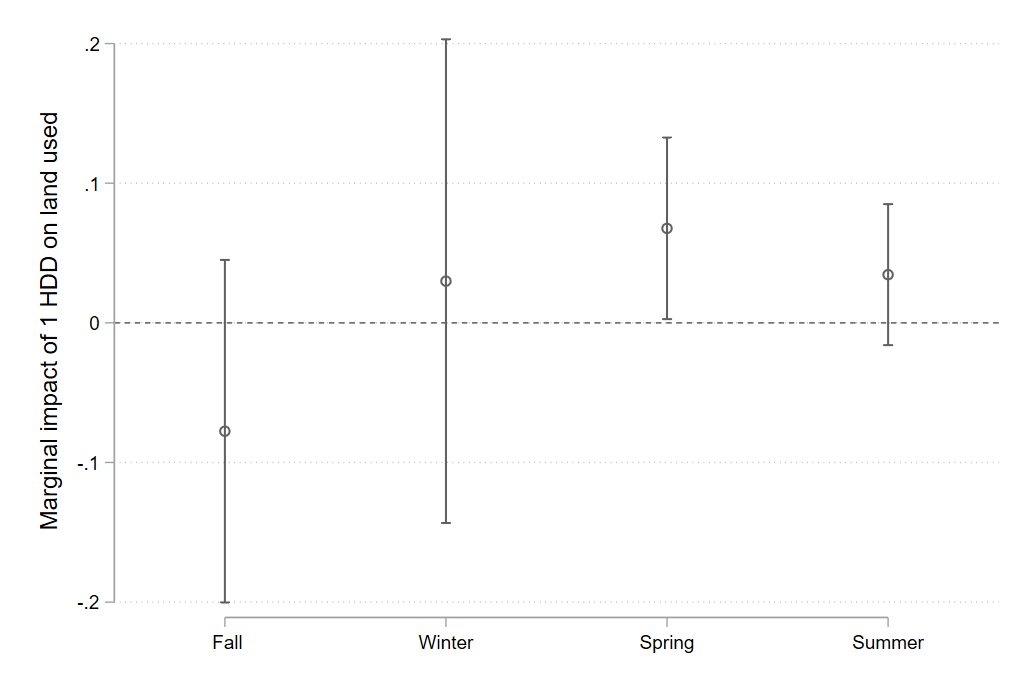}
        \caption{Impacts on ln(hectares planted)}
        \label{fig:timing_T}
    \end{subfigure}
	\vspace{10pt}
	\parbox{12cm}{\footnotesize{\emph{Notes:} points represent the point estimates for HDD-season coefficients from two regression models, one with agricultural yields and the second with total land use as dependent variables (both in logarithms). Lines indicate 95\% confidence intervals. Standard errors clustered at the district level. All specifications include  district, month of interview, and climatic region-by-growing season fixed effects, and the same farmer controls as baseline regression in Table \ref{table_main}. Regression on land used includes the following endowment controls: log of household size and area of land owned.}}
\end{figure}

Figure \ref{fig_timing} depicts the effect of average HDD in different seasons on our measures of productivity (Y/T) and land used (T). The main observation is that the effect of extreme heat on productivity and land use is driven by shocks that occur during the spring.
This timing is consistent with the biological response (and the human reaction) to heat experienced during a sensitive period in the agricultural calendar. Previous studies show that, while plants are vulnerable to high temperatures throughout their life-cycle, the potential harm is highest during the sowing period \citep{Slafer1994}. Moreover, it suggests that the observed changes in land use are a response to productivity shocks within the agricultural season. 


\subsubsection{Changes in labor use}
  
Finally, we examine the effect of extreme heat on labor. We distinguish two types of labor: domestic and hired. In contrast to land use, we do not have good proxies for labor used during the agricultural season. We only observe the wage bill of hired workers in last 12 months, not actual number of workers. More importantly, we only have information on labor outcomes of household members during the last 2 weeks before the interview, not for the whole agricultural year. Because of these limitations, the results on labor use should be interpreted with caution. 

Table \ref{table_labor} presents our findings. Columns 1 to 4 examine the effect on two measures of 
domestic labor: number of household members working on the farm, and an indicator of child labor. We estimate the effect of HDD using the baseline specification (columns 1 and 2) as well as an alternative specification restricting the sample to farmers interviewed in spring and summer and using average HDD in spring as a measure of exposure to extreme heat. By focusing on households interviewed at the moment when most of the productivity shock occurs, we can partially address the data limitations mentioned above. Column 5 examines the effect on wage bill: our proxy for hired labor.

Similar to the results on land used, we find that HDD has a positive and, in most cases, significant effect on measures of domestic labor. Interestingly, extreme heat seems to increase the likelihood of child labor. This last result is consistent with findings in the literature on child labor showing that poor households may resort to employing children in productive activities when subject to negative income shocks \citep{BEEGLE,BANDARA}. In contrast, the coefficient of HHD on hired labor's wage bill is negative, albeit also insignificant. These findings suggest a slight tendency of farms to use more intensively domestic labor as a response to extreme heat.

 \begin{table}[h!]
    	\centering
    	\begin{threeparttable}
    		\caption{Temperature and labor use}
    		\label{table_labor} 
    		\input{content/table_labor.tex}
    		\begin{tablenotes}
    			\footnotesize
    			\item \emph{Notes:} Standard errors clustered at the district level (in parenthesis). Stars indicate statistical significance: \textit{*p \textless 0.10, ** p \textless 0.05, *** p \textless 0.01}. All specifications include  district, month of interview, and climatic region-by-growing season fixed effects, and the same farmer controls as baseline regression in Table \ref{table_main}. Columns 2 and 4 restrict the sample to farmers interviewed during the growing season (spring and summer). Columns 1 and 3 also restrict the sample to households with at least one child aged 6 to 15 years. HH = household.
    		\end{tablenotes}
    	\end{threeparttable}
    \end{table}
    
\subsubsection{Discussion}
Our findings are hard to reconcile with predictions from a standard production model. As discussed in Section \ref{section_model}, a standard production model would predict a weakly negative relation between HDD and input use, as well as a negative effect on output. The reduction in productivity would drive the negative effect on input use. However, if extreme heat shocks occur after input decisions are sunk (i.e., after planting), there would be no effect of HDD on area planted.

Instead, our findings are consistent with models of subsistence farmers in a context of incomplete markets \citep{de1991peasant,taylor2003agricultural}. In this scenario, production and consumption decisions are not separable \citep{Benjamin}. Thus, farmers exposed to negative shocks may need to resort to more intensive use of non-traded inputs, like land and domestic labor, to offset undesirable drops in output and consumption. In this sense, changes in input use are akin to other consumption smoothing mechanisms, such as selling disposable assets or increasing off-farm work \citep{Rosenzweig_Wolpin,Kochar}. 

To the best of our knowledge, this margin of adjustment, namely increasing the intensity of land use,
has not been previously documented in the consumption smoothing literature, nor in existing studies of the effect of temperature on agriculture. However, it may be particularly relevant for farmers in less developed countries due to the presence of several market imperfections and limited coping mechanisms, such as crop insurance or savings. 
 
Moreover, it has at least two important implications. First, it suggests a potential dynamic link between weather shocks and long-run outcomes. Leaving land uncultivated, i.e., fallowing, is a common practice in traditional agriculture to avoid depleting soil nutrients, recover soil biomass, and restore land productivity \citep{goldstein2008profits}.\footnote{In our sample. around 50\% of farmers have some land fallowing.} If the increase in area planted as a response to extreme temperature comes at the expense of fallow land, then this short-term response could affect land productivity in the medium or long term. 
  
Second, this farmer response may affect estimations of the damages of climate change on agricultural output. These estimates are usually based on the effect of temperature on crop yields ($Y/T$). This is a correct approach if land use is fixed. In that case, changes in crop yields are the same as changes in output. However, using crop yields may be less informative in contexts in which farmers respond to weather shocks by changing land use. As we show in Section \ref{section_CC}, taking into account this adaptive response reduces, in a non-trivial magnitude, the predicted damages.

\subsection{Additional checks}\label{section_robustness}

\subsubsection{Alternative specifications}

Table \ref{table_robustness} presents several checks of the robustness of our main results to alternative model specifications. We report only the estimate associated with the measure of extreme heat (HDD). Each row uses a different specification.

Row 1 restricts our sample only to farmers interviewed in fall and winter. By that time, the main growing season has passed and farmers have reaped the main harvest of the year. This specification drops almost half of the baseline sample, but it reduces concerns of measurement error due to mismatch of planting and harvesting decisions, confounding of current and previous weather shocks, or recall bias. Row 2 estimates a more parsimonious model without any individual or household-level controls, only district and region-by-year fixed effects, while row 3 implements a more conservative clustering at province (n= 159) instead of district level (n=977). In all three cases, our results are similar to the baseline specification.

Our results are also robust to alternative ways to measure exposure to extreme heat. Row 4 uses the number of days in growing seasons with HDD, while row 5 uses average HDD during the last 12 months instead of during the last completed growing season. We also obtain similar results when allowing for different HDD thresholds by climatic region, i.e., coast and highlands (row 6).\footnote{These region-specific thresholds were chosen by replicating the analysis shown in Figure \ref{fig_bin_peru} in the coast and highland observations separately. The results from this exercise are presented in Figure \ref{fig:bin_region}, in the Appendix.} Figure \ref{fig:fig_robustness_threshold} in the Appendix further assesses the sensitivity of our results to different values of the threshold ($\tau$) ranging from 26\celsius\ to 42\celsius. These results show that lower thresholds produce similar results, while higher thresholds increase the magnitude of our baseline estimates and reduce their precision.



\subsubsection{Prices as omitted variables}

An important concern is that our results might be driven by changes in relative prices. 
Extreme heat shocks can reduce aggregate supply and increase agricultural prices. This price increase would, in turn, create incentives to increase production and input use. In our baseline specification, we address this concern by including a set of climatic region-by-growing season fixed effects. To the extent that agricultural markets are national or circumscribed to climatic regions, this approach would control for agricultural prices. However, if agricultural markets are narrower, we could have an omitted variables problem.
 
We examine the relevance of this issue in two ways (see rows 7 to 8 in Table \ref{table_robustness}).  First, we add province-by-growing season fixed effects (row 7).  This is a much richer set of time-varying controls than our baseline specification and, under the assumption that agricultural markets are province-wide, effectively controls for prices. Second, we add proxies of local prices at district level (row 8). We focus on tubers and cereals: the two main types of crops in our sample. For each crop type, we construct a price index at the district level and add it to baseline regression.\footnote{The price index for each crop type is a Laspeyres index using self-reported unit prices and output shares of each crop (within a crop group) in baseline year 2007. We then take natural logarithms.} In both cases, our results remain similar to the baseline specification.
 
\begin{table}[h!]
    \centering
    \begin{threeparttable}
        \caption{Robustness checks}
        \label{table_robustness} 
        {
            \def\sym#1{\ifmmode^{#1}\else\(^{#1}\)\fi}
            \begin{tabular}{l*{4}{c}}
                 \input{content/table_robustness.tex}
             \end{tabular}
        }
        \begin{tablenotes}
            \footnotesize
            \item \emph{Notes:} Standard errors clustered at the district level (in parenthesis). Stars indicate statistical significance: \textit{*p \textless 0.10, ** p \textless 0.05, *** p \textless 0.01}. All specifications, except in row 2 include the same controls as baseline regression in Table \ref{table_main}. Row 1 restricts the sample to farmers interviewed in fall and winter (i.e., April to August). Row 7 adds province-by-growing season fixed effects while row 8 includes logs of price indexes for tubers and cereals at district level. 
        \end{tablenotes}
    \end{threeparttable}
\end{table}

\subsubsection{Regional differences}
   
As discussed in Section \ref{section_subsistence}, our sample has two distinct climatic regions: coast and highlands. The coast has a warm semi-arid climate with very little precipitation, especially in the central and southern coast. In contrast, the highlands are cooler and receive more rain. These climatic differences are apparent when observing the distribution of daily temperature in these two regions (see Figure \ref{fig:bin_region} in the Appendix). The two regions also differ in their agricultural practices. Coastal farmers are, on average, substantially better off, are more productive, more educated, and more likely to have access to irrigation. Compared to highland farmers, coastal farmers are also more likely to specialize on fruits and cereals, less likely to own livestock and cultivate a larger share of their land. 

Given these regional differences, a relevant question is whether our baseline specification, which pools all observations, may be hiding relevant heterogeneity in the effects and responses to extreme heat. We address this question by relaxing the baseline specification and allowing for different 
effects of weather variables (DD, HDD, and precipitation) by climatic region.\footnote{We modify the baseline specification by including interaction terms of weather variables with an indicator of being located in the highlands.} Table \ref{table_region} shows the estimates of the effect of HDD for each region, and displays the p-value of the test of equality of both estimates. 

Our main conclusions still remain the same after allowing for regional differences: in both regions, extreme heat has a negative effect on productivity and a positive effect on the quantity of land used. 
Surprisingly, despite coastal farmers being normally exposed to higher temperatures, the magnitude of the effect on yields is similar in both regions. This result echoes findings by \citet{BurkeEmerick} among U.S. corn farmers. Using a long difference approach, they find that extreme heat has similar detrimental effects on crop yields across time, despite the observed increase in average temperatures. \citet{BurkeEmerick} interpret this finding as suggestive evidence of limited long-term adaptation to higher temperatures.

There are, however, some quantitative differences on the effect on land use. In particular, the increase in area planted is smaller in the coast. In this region, there is also no significant change in crop mix, measured by the share of tubers in total output. There are several possible explanations for these differences. First, they may reflect lower land availability in the coast. In this region, agriculture occurs in densely populated valleys, surrounded by very arid deserts, and depends heavily on access to irrigation.\footnote{The share of uncultivated land is almost 45\% in the highlands and 11.5\% in the coast (see Table \ref{table_summary}).}
These features can constrain the expansion of agricultural land. 
Second, they may be driven by coastal farmers having access to other coping mechanisms. This is plausible given that coastal farmers tend to be better off and are closer to cities and other urban areas. Finally, we cannot rule out that these results are driven by low statistical power given the relatively small sample of coastal farmers (around 13\% of our sample).  
 
  \begin{table}[h!]
    	\centering
    	\begin{threeparttable}
    		\caption{Effect of HDD on productivity and land use, by climatic region}
    		\label{table_region} 
    		\input{content/table_region.tex}
    		\begin{tablenotes}
    			\footnotesize
    			\item \emph{Notes:} Standard errors clustered at the district level (in parenthesis). Stars indicate statistical significance: \textit{*p \textless 0.10, ** p \textless 0.05, *** p \textless 0.01}. All specifications include  district, month of interview, and climatic region-by-growing season fixed effects, and the same farmer controls as baseline regression in Table \ref{table_main}. 
    		\end{tablenotes}
    	\end{threeparttable}
    \end{table}

\section{Other coping mechanisms}\label{section_other_coping}

Our main results suggest that farmers adjust input use as a mechanism to cope with the negative effects of extreme temperatures. In this section, we study other coping mechanisms previously documented in the consumption smoothing literature, such as working in non-agricultural activities \citep{Rosenzweig_Stark,Kochar,colmer2016weather}, migrating \citep{Munshi,feng2012climate,Kleemans_Magruder,jessoe2017climate} or selling livestock \citep{Rosenzweig_Wolpin}. Then, we examine how these coping mechanisms interact with changes in land use. 

We start by examining whether farmers in our context use other coping mechanisms (see Table \ref{table_other}). Our first set of outcomes focuses on the use of livestock as a buffer against income shocks (columns 1 to 3). We find that HDD is associated with an increase in the probability that a farmer reports a decrease in livestock value.\footnote{Our definition of livestock includes cattle,  horses, sheep, llamas, and pigs.} This reduction seems to come from households selling, rather than consuming their livestock. These results are consistent with farmers selling livestock to offset the adverse effects of extreme heat.

Next, we focus on indicators of off-farm work (columns 4 and 5). We use an indicator of a household member having a non-agricultural job, as well as the total number of hours worked off-farm (conditional on having a non-agricultural job). As in Table \ref{table_labor}, we restrict the sample to households interviewed during the growing season (i.e., spring and summer). These outcomes capture supply of off-farm employment in the extensive and intensive margin. In the extensive margin, the estimate is insignificant. However, the estimate on the intensive margin is positive and statistically significant: farmers with off-farm jobs seem to increase the number of hours worked in that activity. While suggestive of off-farm employment as a coping strategy, this result is not robust to using the whole sample of farmers. 

In columns 6 and 7, we look for evidence of short-term migration. Due to data limitations, we cannot measure migration directly. Instead, we use proxy variables such as an indicator of whether any member has been away for more than 30 days and household size. Similar to the results on off-farm employment, none of these outcomes seems to be affected by extreme temperature. However, we should interpret these last results with caution. Our analysis focuses on a short period (within a year), and these adjustments may happen over a longer time frame. In addition, our measures of labor and migration may be noisy proxies of actual behavior. These factors likely reduce the power of our statistical analysis and could explain the insignificant results.
 

  \begin{landscape}
  \begin{table}[h!]
        \centering
        \begin{threeparttable}
            \caption{Other responses to extreme heat}
            \label{table_other} 
            \input{content/table_other.tex}
            \begin{tablenotes}
                \footnotesize
                \item \emph{Notes:} Standard errors clustered at the district level (in parenthesis). Stars indicate statistical significance: \textit{*p \textless 0.10, ** p \textless 0.05, *** p \textless 0.01}. All specifications include  district, month of interview, and climatic region-by-growing season fixed effects, and the same farmer controls as baseline regression in Table \ref{table_main}. Columns 1 to 3 restrict the sample to farmers who reported having livestock 12 months ago. Columns 4 and 5 restrict the sample to farmers interviewed in spring and summer. Column 5 further restricts the sample to households in which at least one member has an off-farm job. All regressions are estimated using OLS. All regressions, except in columns 5 and 7, have a binary outcome variable.
            \end{tablenotes}
        \end{threeparttable}
    \end{table} 
  \end{landscape}

\subsection{Interactions with productive responses}

Our results suggest that, in our sample, farmers seem to use livestock sales as a coping strategy to smooth negative weather shocks. A natural question is how this coping strategy interacts with the productive responses, such as increasing input use, identified in our main results. Does having livestock eliminate the need to change land use, or do they complement each other? These are relevant questions to better understand the portfolio of coping strategies available to subsistence farmers. 

We examine these issues by estimating heterogeneous responses to extreme heat for farmers with different ability to use other coping strategies. Based on our previous findings, we interact HDD with indicators of owning livestock 12 months ago and having at least one household member employed in a non-agricultural activity. We use these indicators as proxies of farmers' ability to use livestock and off-farm employment as buffers to negative income shocks.

Our results in Table \ref{table_who_adapts} suggest that the effect of HDD on land use (area planted and relative share of tubers) is qualitatively similar between farmers with and without livestock (columns 2 and 3). However, the magnitude of the effect is larger among farmers who do not own livestock. This result is not driven by these latter farmers experiencing a larger negative productivity shock. As shown in column 1, the effect of HDD on agricultural yields is similar for both types of farmers and, if anything marginally smaller for farmers without livestock. In the case of off-farm employment (columns 4 to 6), there are no significant quantitative differences in the effect of HDD in any outcome.

We interpret these results as evidence that farmers do not use one strategy exclusively but instead use a combination of responses to cope with extreme heat. These responses include both sale of disposable assets (such as livestock) and adjustments in production decisions (such as changes in land use). 

\begin{table}[h!]
        \centering
        \begin{threeparttable}
            \caption{Interaction with changes in land use}
            \label{table_who_adapts} 
            \input{content/table_who_adapts.tex}
            \begin{tablenotes}
                \footnotesize
                \item \emph{Notes:} Standard errors clustered at the district level (in parenthesis). Stars indicate statistical significance: \textit{*p \textless 0.10, ** p \textless 0.05, *** p \textless 0.01}. All specifications include  district, month of interview, and climatic region-by-growing season fixed effects, and the same farmer controls as baseline regressions in Table \ref{table_main}. 
                Regressions includes interaction of HDD with an indicator variable $D$ of whether household owned livestock 12 month ago  (columns 1 to 3) or has a member with a non-agricultural job (columns 4 to 6).  All regressions also include the interaction of HDD with an indicator of climatic region. The third row reports the p-value of test of equality of estimates in first two rows. 
            \end{tablenotes}
        \end{threeparttable}
    \end{table} 

\section{Implications for estimating damages from climate change}\label{section_CC}
 
Most models assessing climate change damages use estimates of the effect of temperature on crop yields to calculate the loss of agricultural output and, hence, rural income. This approach is correct if, among other things, the amount of land used is constant. However, if farmers increase land use, as we have documented above, this approach would ignore an important margin of productive adaptation and overestimate the actual fall in agricultural output.

In this section, we quantitatively assess the magnitude of this overestimation of damages. 
To do so, we obtain end-of-the-century predictions of temperature over our study area from current climate change projections. Then, we calculate the predicted change in agricultural output by extrapolating the effect of these temperatures on agricultural yields. This is the approach commonly used in the literature.\footnote{See, for example, \cite{Deschenes2007}} Finally, we compare these results to predictions obtained using our estimates of the effect of temperature on output. These latter estimates take into account changes in land use.

Importantly, this exercise only assumes changes in temperature (DD and HDD) and keeps everything else constant. Thus, it does not account for other potential factors and responses associated with climate change such as changes in CO$_2$, increase risk of natural disasters, changes in water availability, degradation of land quality, migration, changes in sectoral employment, etc. For that reason, our results should be interpreted with caution: they do not attempt to predict the effect of global warming on Peruvian agriculture, but only to highlight the importance of accounting for farmers' changes of land use when estimating damages from climate change.

\subsection{Climate change projections} 

We obtain temperature projections from two climate change scenarios: RCP45 and RCP85. These scenarios, used in the IPCC's Fifth Assessment Report \citep{ipccsynth}, represent two different sets of assumptions about the future trajectory of global greenhouse gas emissions.\footnote{We use the model output produced by the Hadley Centre Global Environment Model version 2 (HadGEM2-ES).} RCP85 is a `business as usual' framework in which no additional policies to reduce greenhouse gas emissions are introduced. This scenario forecasts an increase of 4.9 \celsius\ in global temperatures by the end of the century. RCP45 is a more optimistic scenario that assumes increased efforts to curb emissions at a global scale and forecasts an average 2.4 \celsius\ increase in global temperatures. 
 
For each scenario, we obtain gridded data at a resolution 1.25 x 1.875 degrees of monthly temperatures for the baseline year 2005 and the forecast for the year 2099. We then adjust for model-specific error in a similar way to \cite{deschenes2011} to account for the fact that the historical temperatures (from MODIS) and predicted temperatures (from the HadGEM2-ES model) are from different sources.\footnote{We calculate the implied temperature change (i.e., 2099 compared to 2005) for each month-location according to each HadGEM2 scenario, and then add this to the average temperature in our (MODIS) dataset for each day of the year.} Then, we use the predicted temperature distribution for each scenario $j$ and location $k$ to calculate $DD_{jk}$ and $HDD_{jk}$ for the end of the century. \footnote{We assume the same optimal temperature threshold as discussed in the previous section, 33\celsius. In both scenarios, average precipitation is predicted to stay within one standard deviation of its natural internal variability, so we do not assume any change in this respect \citep{ipccsynth}.}
 
Panel A in Table \ref{table_CC} presents the predicted average $\Delta DD$ and $\Delta HDD$ for our whole sample and each climatic region in both scenarios.\footnote{Formally, $\Delta HDD_{jk} =HDD_{k}  - \bar{HDD}_{k}$ where  $\bar{HDD}_{k}$ is the average historical  HDD in  location $j$. We use a similar procedure to calculate the change in degree-days $\Delta DD_{jk}$.} Note that the increase in average HDD is 0.639 \celsius\ in the RCP45 scenario and more than double, 1.323 \celsius\, in the `business as usual' scenario. The increase in temperature will create substantially more harmful temperatures in the coast than in the highlands. While the coast is expected to experience 3-5 additional harmful degrees a day during growing season months, the highlands are expected to experience just up to 0.7 HDD a day, in the most pessimistic scenario. These results are a natural consequence of the current distribution of temperatures in both regions: as previously mentioned, the coast is already drier and hotter than the highlands. Thus, a shift in the distribution of temperature has a larger effect on the frequency of extremely hot days.
 
\subsection{Predicted effects on agriculture} 
 
We calculate the predicted change on agricultural yields and output using the estimated effect of temperature on agricultural outcomes and the predicted changes in temperatures from climate change forecasts. In particular, we calculate the predicted effects as follows:

\begin{eqnarray*}
   \Delta y_{ijk} = \hat{\beta_{1}} \Delta DD_{jk}  + \hat{\beta_{2}} \Delta HDD_{jk}  
\end{eqnarray*}

where $y$ is the outcome (i.e,  yield or output) of farmer $i$ in location $k$, while $\hat{\beta_{1}}$ and $\hat{\beta_{2}}$ correspond to the estimated effect of DD and HDD for each climatic region (coast and highlands) taken from columns 1 and 2 in Table \ref{table_region}.
 
Panels B and C in Table \ref{table_CC} present our results. The main observation is that using yields to predict the effect of climate change can lead to a substantial \textit{overestimation} of the loss of agricultural output. This finding suggests that taking into account farmers' adjustments in land use is quantitatively important when estimating damages associated with climate change. 

For instance, assuming the quantity of land used is fixed, we would predict that drops in output are equal to drop in yields. This implies a drop in output of around 5 to 9 percent (columns 1 and 4). However, the predicted drop in output is much smaller: around 0.6 to 1.2 percent. Overestimation is particularly salient in the coast. In that region, assuming land used is fixed, output losses are estimated to range from 29 to 48 percent. These magnitudes are almost twice as large as when allowing for changes in land used. In the highlands, the differences when using both types of approaches are much smaller, but they produce qualitatively different results: a drop in yields, but an increase in output.

Naturally, land is a finite resource, and thus this particular strategy is not dynamically consistent. In other words, farmers will not be able to offset output losses in the face of higher temperatures by adding more land to their production function indefinitely. Nevertheless, note that the farmers in our sample keep large amounts of unused land during any given growing season (see Table \ref{table_summary}). In the case of highland farmers this is as high as 40\% of their land holdings. It is, therefore, a productive adaptation with a significant margin over the near term.  

As a final point, our predictions highlight potentially heterogeneous impacts on agricultural production: while the coast will experience sizable output losses, the impact in the highlands would be slightly positive. This result is consistent with other studies that predict large negative effects of climate change on low-lying areas, and milder, even positive, effects in cooler areas \citep{Deschenes2007,auffhammer2014empirical}.

   \begin{table}[h!]
        \centering
        \begin{threeparttable}
            \caption{Predicted effects of climate change on agriculture}
            \label{table_CC} 
            \input{content/table_CC.tex}
        \end{threeparttable}
    \end{table} 
    

\section{Conclusion}\label{section_conclusion}

This paper examines how subsistence farmers respond to extreme temperature. Using micro-data from Peruvian farmers, we show that extreme heat events decrease agricultural productivity, but increase area planted. The expansion of area planted is coupled with changes in crop mix. We also find suggestive evidence of an increase in domestic labor.

We interpret these results as evidence that farmers use productive adjustments, such as changes in input use, as strategies to attenuate drops in output and consumption. This interpretation is consistent with predictions of producer-consumer models in the presence of incomplete markets.
 
Our results highlight a margin of adjustment not previously documented in the literature. We argue that this response may be relevant to subsistence, traditional, farmers in other developing countries. These type of farmers constitute a large fraction of agricultural workers and the rural poor. However, the study of how extreme temperatures affect them has been so far neglected.

Due to data limitations, some important questions raised in this paper remain unanswered. First, we cannot observe the exact timing of planting, only the total amount of land used. Thus, we are unable to examine if the land response is an ex-ante or ex-post adaptation to weather shocks.  Second, we only observe farmers for a year. This limited period reduces our ability to study long-term adaptive strategies, such as permanent migration,  changes in agricultural technology, or the potential dynamic effects of short term responses, such as land degradation or loss of human capital. Finally, common to other recent studies of the climate-economy literature, we can only observe the effect of weather shocks, not of climatic changes. 


\newpage

\bibliography{climate}
	\bibliographystyle{aer}


\newpage
\appendix
      
 \input{appendix}

\end{document}

%% file: content/table_summary.tex
{
\def\sym#1{\ifmmode^{#1}\else\(^{#1}\)\fi}
\begin{tabular}{l*{3}{c}}
\toprule
                    &\multicolumn{1}{c}{(1)}&\multicolumn{1}{c}{(2)}&\multicolumn{1}{c}{(3)}\\
                    &\multicolumn{1}{c}{All}&\multicolumn{1}{c}{Coast}&\multicolumn{1}{c}{Highlands}\\
\midrule
\emph{A. Household characteristics} \rule{0pt}{4ex}&            &            &            \\
\hspace{0.1cm} Poor (\%)&       51.14&       26.55&       55.10\\
\hspace{0.1cm} Household size&        4.34&        4.41&        4.33\\
\hspace{0.1cm} Primary education completed by HH head (\%)&       50.93&       58.48&       49.71\\
\hspace{0.1cm} Child works (\%)&       21.82&        9.65&       23.79\\
\hspace{0.1cm} At least 1 HH member has off-farm job (\%)&       47.54&       56.45&       46.10\\
\emph{B. Agricultural characteristics} \rule{0pt}{4ex}&            &            &            \\
\hspace{0.1cm} Value of agric. output (Y) &     1049.93&     3263.23&      693.40\\
\hspace{0.1cm} Output per ha. (Y/T) &     1048.92&     1868.49&      917.09\\
\hspace{0.1cm} Land used (T), in ha.&        1.99&        2.41&        1.92\\
\hspace{0.1cm} No. HH members work on-farm&        2.31&        2.21&        2.33\\
\hspace{0.1cm} Hire workers (\%)&       48.85&       57.08&       47.52\\
\hspace{0.1cm} Uncultivated land (\% of land holding)&       40.30&       11.81&       44.89\\
\hspace{0.1cm} Irrigated land (\% land holding)&       36.05&       82.00&       28.65\\
\hspace{0.1cm} Fruits (\% total output)&        7.41&       31.59&        3.52\\
\hspace{0.1cm} Tubers (\% total output)&       31.35&        5.54&       35.50\\
\hspace{0.1cm} Cereals (\% total output)&       31.30&       30.43&       31.44\\
\hspace{0.1cm} Own livestock (\%)&       77.61&       55.95&       81.10\\
\hspace{0.1cm} Value of livestock &      682.11&      461.85&      717.59\\
\emph{C. Weather during the last growing season} \rule{0pt}{4ex}&            &            &            \\
\hspace{0.1cm} Average temperature (\celsius)&       22.84&       33.07&       21.20\\
\hspace{0.1cm} Average DD&       14.28&       22.39&       12.97\\
\hspace{0.1cm} Average HDD&        0.73&        2.69&        0.41\\
\hspace{0.1cm} \% days with HDD&        0.17&        0.53&        0.11\\
\hspace{0.1cm} Precipitation (mm/day)&        3.16&        0.93&        3.51\\ [1ex]
Observations&      53,619&       7,439&      46,180\\
\bottomrule
\end{tabular}
}

%% file: content/table_other.tex
{
\def\sym#1{\ifmmode^{#1}\else\(^{#1}\)\fi}
\begin{tabular}{l*{9}{c}}
 
    \toprule
          & \multicolumn{3}{c}{Livestock buffer} & \phantom{}      & \multicolumn{2}{c}{Off-farm work} &   \phantom{}     & \multicolumn{2}{c}{Short-term migration} \\
\cmidrule{2-4}\cmidrule{6-7}\cmidrule{9-10}    Dep. variable: & Decrease in & Sold  & Consumed &       & HH member  & ln(hours &       & HH member &  \\
          & livestock & livestock & livestock &       & has off-farm & worked &       & away 30+ & HH size \\
          & value &       &       &       & job   & off-farm) &       & days  &  \\
          & (1)   & (2)   & (3)   &       & (4)   & (5)   &       & (6)   & (7) \\
    \midrule
          &       &       &       &       &       &       &       &       &  \\
    Average DD & -0.008*** & -0.012*** & -0.013*** &       & 0.009** & 0.026*** &       & 0.003** & -0.006 \\
    & (0.002) & (0.002) & (0.003) &       & (0.004) & (0.009) &       & (0.001) & (0.014) \\
          &       &       &       &       &       &       &       &       &  \\
    Average HDD & 0.022*** & 0.016* & 0.007 &       & 0.006 & 0.054** &       & -0.002 & 0.016 \\
    & (0.007) & (0.009) & (0.009) &       & (0.011) & (0.025) &       & (0.002) & (0.033) \\
          &       &       &       &       &       &       &       &       &  \\
    Mean outcome & 0.332 & 0.517 & 0.476 &       & 0.464 & 57.548 &       & 0.085 & 4.339 \\
          &       &       &       &       &       &       &       &       &  \\
    No. obs. & 48,169 & 48,169 & 48,169 &       & 26,726 & 12,377 &       & 53,619 & 53,619 \\
    R-squared & 0.077 & 0.146 & 0.240 &       & 0.213 & 0.169 &       & 0.083 & 0.244 \\
    \bottomrule
\end{tabular}
}

%% file: content/table_who_adapts.tex
{
\def\sym#1{\ifmmode^{#1}\else\(^{#1}\)\fi}
\begin{tabular}{l*{7}{c}}

      \toprule
          & \multicolumn{3}{c}{Livestock buffer} &   \phantom{}    & \multicolumn{3}{c}{Off-farm work} \\
\cmidrule{2-4}\cmidrule{6-8}          & ln(output & ln(area  & Tubers &       & ln(output & ln(area  & Tubers \\
    Dep. variable:  & per ha) & planted) & \% output &       & per ha) & planted) & \% output \\
          & (1)   & (2)   & (3)   &       & (4)   & (5)   & (6) \\
    \midrule
          &       &       &       &       &       &       &  \\
    (A)  Average HDD $\times$ & -0.087* & 0.055*** & 0.024*** &       & -0.100** & 0.041** & 0.017*** \\
    $D=0$   & (0.044) & (0.019) & (0.005) &       & (0.048) & (0.018) & (0.005) \\
          &       &       &       &       &       &       &  \\
    (B)  Average HDD $\times$ & -0.121** & 0.018 & 0.015*** &       & -0.112** & 0.040** & 0.020*** \\
    $D=1$   & (0.047) & (0.018) & (0.005) &       & (0.046) & (0.018) & (0.005) \\
          &       &       &       &       &       &       &  \\
    Diff. (B)-(A) & 0.059 & 0.002 & 0.000 &       & 0.392 & 0.956 & 0.113 \\
    p-value &       &       &       &       &       &       &  \\
          &       &       &       &       &       &       &  \\
    $D$ is indicator $=1$ if & \multicolumn{3}{c}{HH owns livestock} &       & \multicolumn{3}{c}{Any HH member has off-farm job} \\
          &       &       &       &       &       &       &  \\
    No. obs. & 53,493 & 53,493 & 53,619 &       & 53,493 & 53,493 & 53,619 \\
    R-squared & 0.336 & 0.452 & 0.525 &       & 0.335 & 0.444 & 0.525 \\
    \bottomrule
\end{tabular}
}

%% file: appendix.tex
\begin{center}
   \textbf{ \LARGE{ONLINE APPENDIX - Not for publication }}
\end{center}

\counterwithin{figure}{section}
\counterwithin{table}{section}


    
\section{Additional Figures}

 \begin{figure}[h]
	\centering
	\caption{ENAHO observations 2007-2015} \label{fig_mapPeru}
	\includegraphics[width=13cm]{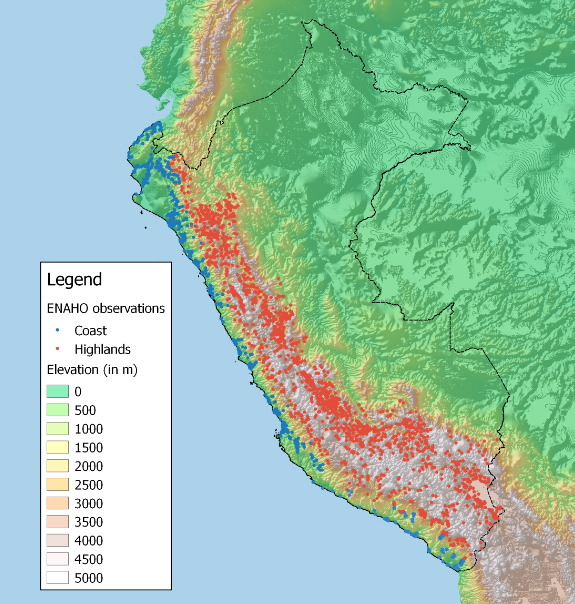}
   	\vspace{20pt}
   	\parbox{13cm}{\footnotesize{\emph{Notes:} Map depicts Peru's climatic regions and location of the 
   	ENAHO clusters used in this study.}}
\end{figure}

\begin{figure}[h]
   	\centering
   	\caption{Distribution of daily average temperature by growing season} \label{fig_distbygs}
   	\includegraphics[width=15cm]{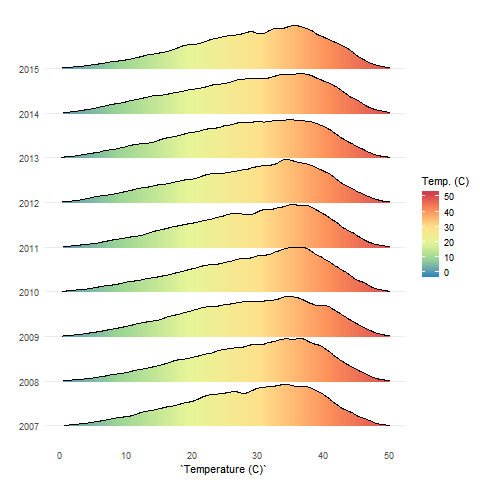}
   	\footnotesize
	\vspace{10pt}
   	\parbox{15cm}{\footnotesize{Notes: Figure depicts the share of days spent in each temperature bin by the farmers in our sample, during the 2007-2015 growing seasons (i.e., October to March).}}
\end{figure}

\begin{figure}[h!]
    \centering
    \caption{Optimal temperature threshold using the iterative regression approach}\label{fig:threshold_test}
    \includegraphics[width=0.9\textwidth]{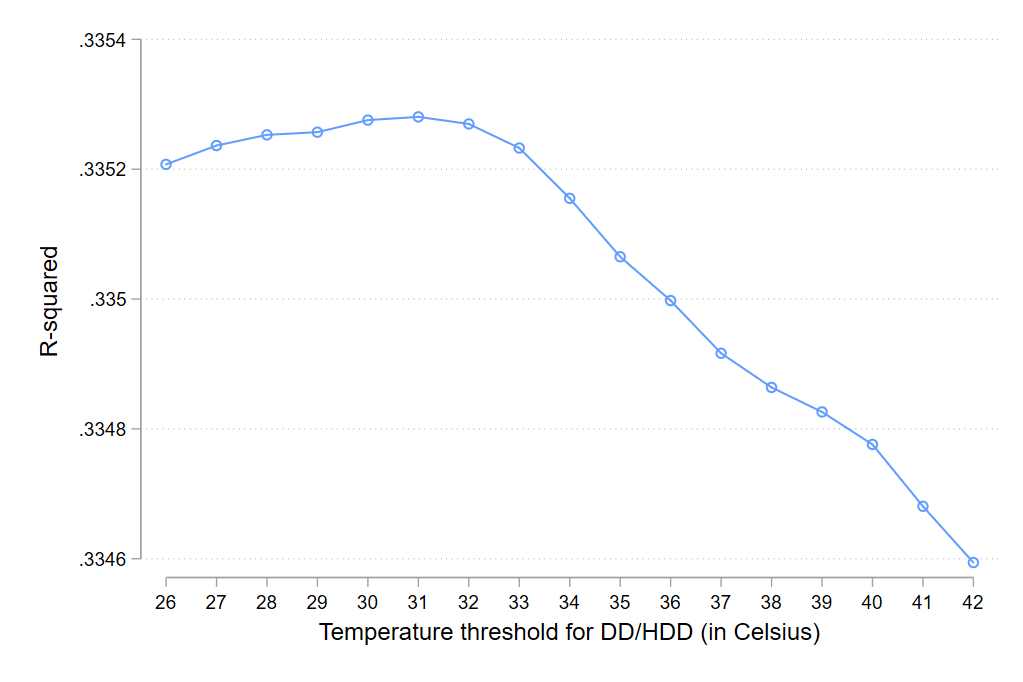}
\end{figure}

\begin{figure}[h!]
    \centering
    \caption{Effect of HDD on yields and land use using alternative DD/HDD thresholds}\label{fig:fig_robustness_threshold}
    \begin{subfigure}[b]{0.7\textwidth}
        \includegraphics[width=\textwidth]{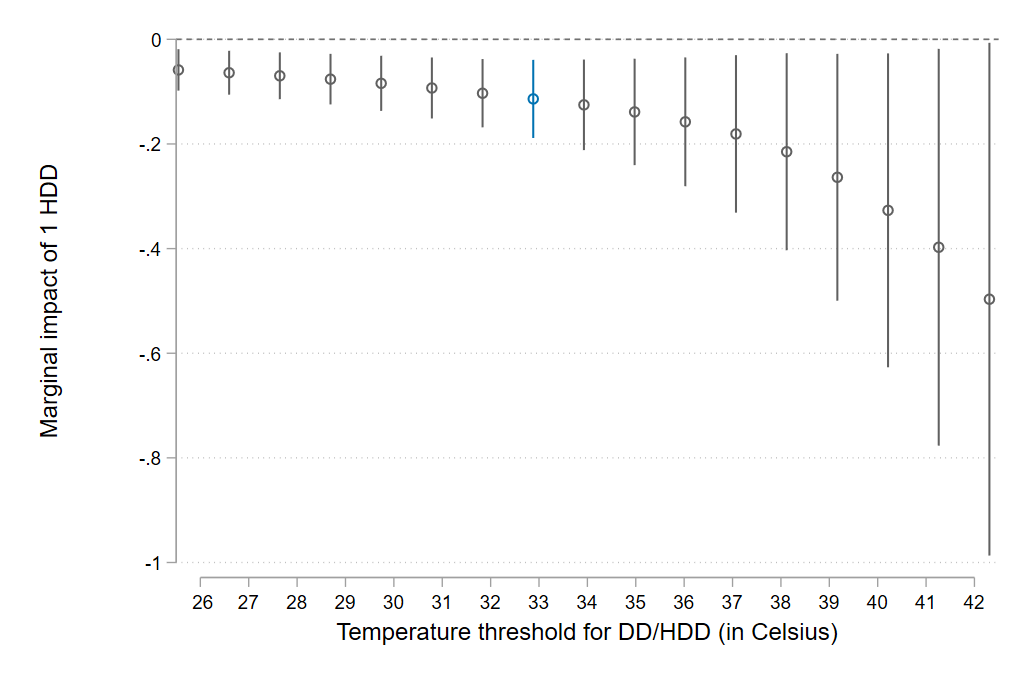}
        \caption{Impacts on ln(output per hectare planted)}
        \label{fig:fig_robustness_threshold1}
    \end{subfigure}
    ~ 
    \begin{subfigure}[b]{0.7\textwidth}
        \includegraphics[width=\textwidth]{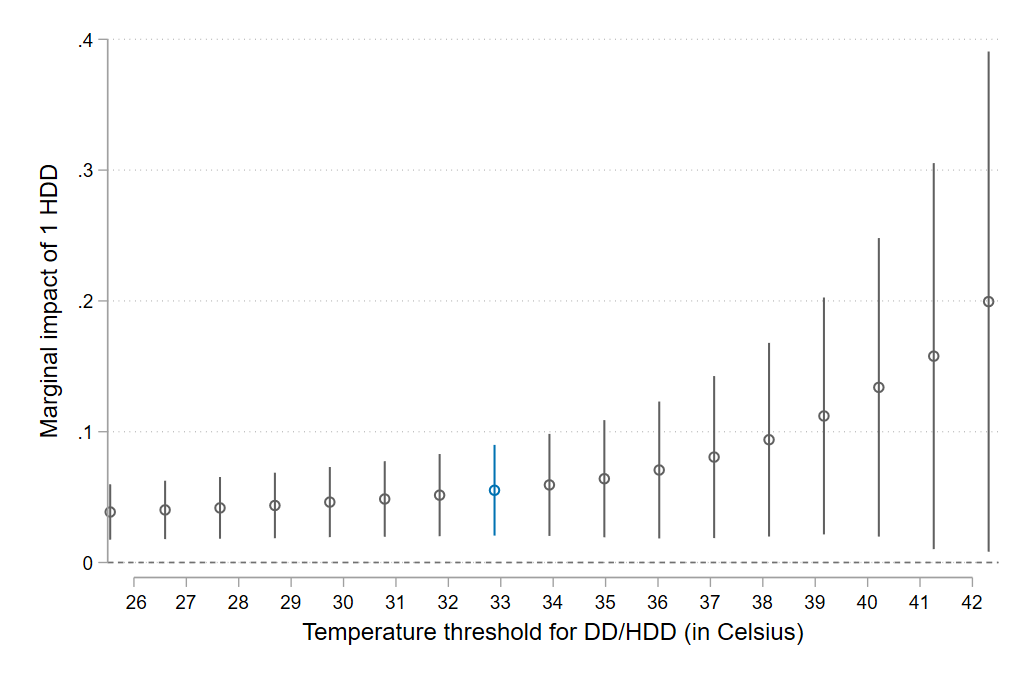}
        \caption{Impacts on ln(hectares planted)}
        \label{fig:fig_robustness_threshold2}
    \end{subfigure}
\end{figure}

\begin{figure}[h!]
    \centering
    \caption{Non-linear relationship between temperature and agricultural yields by region}\label{fig:bin_region}
    \begin{subfigure}[b]{0.7\textwidth}
        \includegraphics[width=\textwidth]{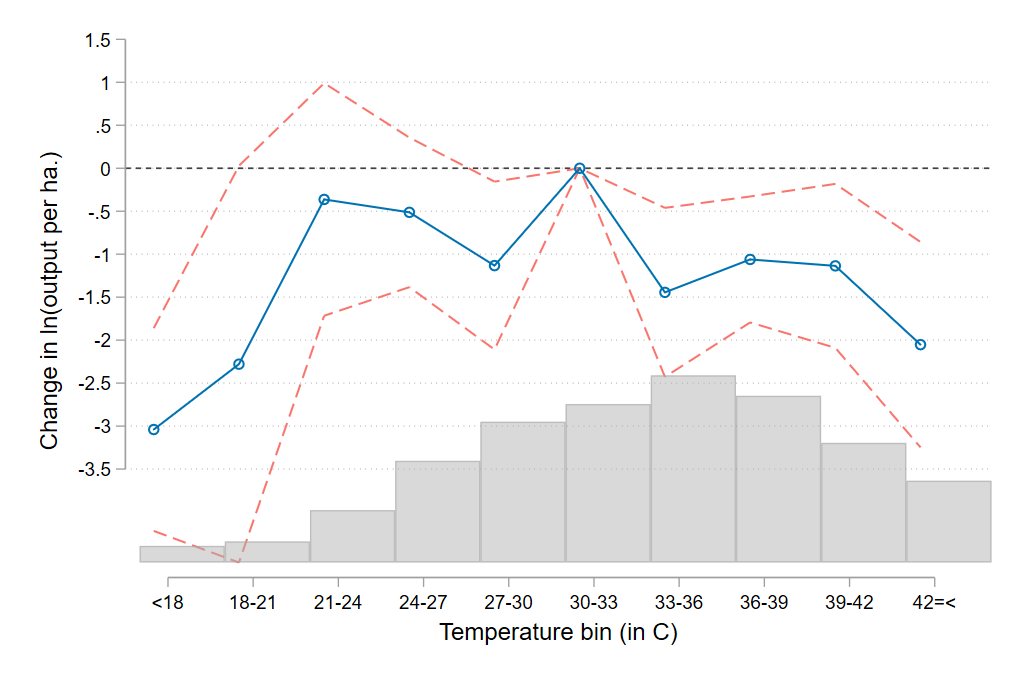}
        \caption{Coast}
        \label{fig:bin_coast}
    \end{subfigure}
    ~ 
    \begin{subfigure}[b]{0.7\textwidth}
        \includegraphics[width=\textwidth]{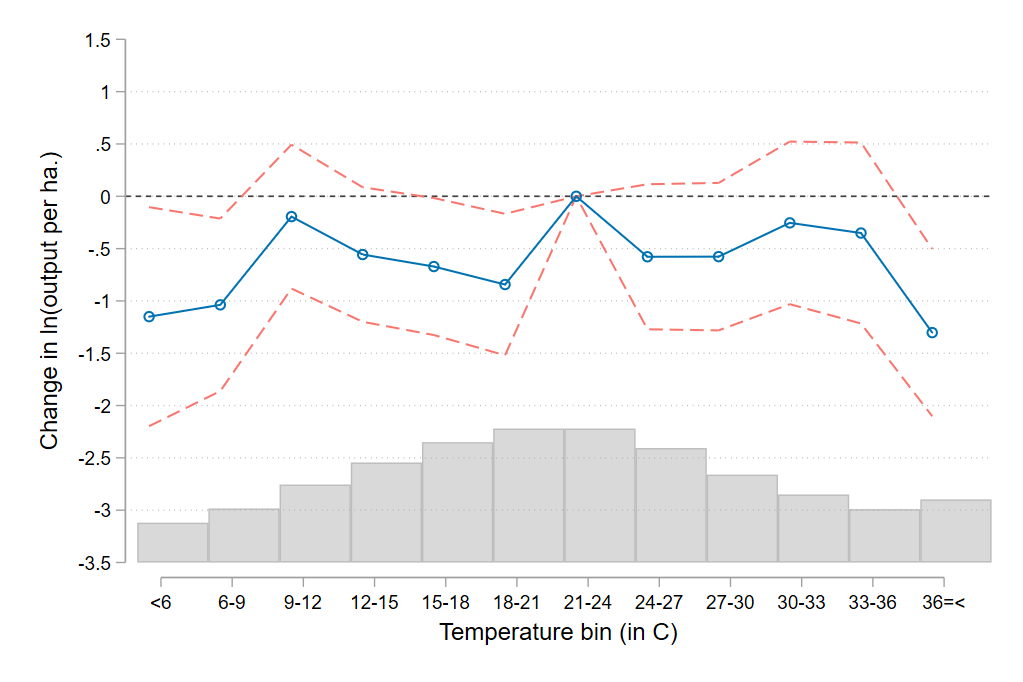}
        \caption{Highlands}
        \label{fig:bin_highlands}
    \end{subfigure}
\end{figure}





\section{Additional tables}

\begin{table}[h!]
	\centering
	\begin{threeparttable}
		\caption{Effect of HDD on other farm inputs}
		\label{t:farminputs} 
		\input{content/Table_otherinputs.tex}
		\begin{tablenotes}
			\footnotesize
			\item \emph{Notes:} Extensive margin use is studied using a dummy variable equal to one if the farmer reports to have used fertilizers/pesticides during the last growing season. Intensive margin use is defined as the logarithm of total amounts spent on fertilizers/pesticides. Standard errors clustered at the district level (in parenthesis). Stars indicate statistical significance: \textit{*p \textless 0.10, ** p \textless 0.05, *** p \textless 0.01}. All specifications include  district, month of interview, and climatic region-by-growing season fixed effects, and the same farmer controls as baseline regression in Table \ref{table_main}.   
		\end{tablenotes}
	\end{threeparttable}
\end{table}
 
 \begin{table}[h!]
 	\centering
 	\begin{threeparttable}
 		\caption{Effect of temperature on farm labor inputs, by type of farmer}
 		\label{t:lab_type} 
 		\input{content/Table_Interactions_lab.tex}
 		\begin{tablenotes}
 			\footnotesize
 			\item \emph{Notes:} Standard errors clustered at the district level (in parenthesis). Stars indicate statistical significance: \textit{*p \textless 0.10, ** p \textless 0.05, *** p \textless 0.01}. All specifications include  district, month of interview, and climatic region-by-growing season fixed effects, and the same farmer controls as baseline regression in Table \ref{table_main}. Sample restricted to interviews conducted during the growing season (i.e. October to March) in columns 1 and 2, since dependent variable is defined as work conducted over the past week. In column 3, we restrict the sample to households with children between the ages of 6 and 15. 
 		\end{tablenotes}
 	\end{threeparttable}
 \end{table}
 
 \begin{landscape}
 \begin{table}[h!]
 	\centering
 	\begin{threeparttable}
 		\caption{Effect of temperature on household income, consumption and poverty rates}
 		\label{table_consumption} 
 		\input{content/Table_consumption.tex}
 		\begin{tablenotes}
 			\footnotesize
 			\item \emph{Notes:} Standard errors clustered at the district level (in parenthesis). Stars indicate statistical significance: \textit{*p \textless 0.10, ** p \textless 0.05, *** p \textless 0.01}. All specifications include  district, month of interview, and climatic region-by-growing season fixed effects, and the same farmer controls as baseline regression in Table \ref{table_main}.
 		\end{tablenotes}
 	\end{threeparttable}
 \end{table}
 
 \end{landscape}

%% file: content/Table_otherinputs.tex
{
\def\sym#1{\ifmmode^{#1}\else\(^{#1}\)\fi}
\begin{tabular}{l*{4}{c}}
\toprule
                    &\multicolumn{2}{c}{Fertilizers}            &\multicolumn{2}{c}{Pesticides}             \\\cmidrule(lr){2-3}\cmidrule(lr){4-5}
                    &\multicolumn{1}{c}{(1)}         &\multicolumn{1}{c}{(2)}         &\multicolumn{1}{c}{(3)}         &\multicolumn{1}{c}{(4)}         \\
Dep var:            &   Extensive         &   Intensive         &   Extensive         &   Intensive         \\
\midrule
\hspace{0.1cm} Average DD&      -0.003         &      -0.021         &       0.001         &       0.002         \\
                    &     (0.003)         &     (0.022)         &     (0.004)         &     (0.018)         \\
\addlinespace
\hspace{0.1cm} Average HDD&       0.003         &       0.002         &       0.005         &       0.029         \\
                    &     (0.010)         &     (0.052)         &     (0.008)         &     (0.043)         \\
\midrule
No. obs.            &      53,619         &      53,618         &      53,619         &      53,618         \\
R-squared           &       0.272         &       0.375         &       0.245         &       0.354         \\
\bottomrule
\end{tabular}
}

%% file: content/Table_Interactions_lab.tex
{
\def\sym#1{\ifmmode^{#1}\else\(^{#1}\)\fi}
\begin{tabular}{l*{4}{c}}
\toprule
                    &\multicolumn{3}{c}{Household Labor}                              &\multicolumn{1}{c}{Hired Labor}\\\cmidrule(lr){2-4}\cmidrule(lr){5-5}
                    &\multicolumn{1}{c}{(1)}         &\multicolumn{1}{c}{(2)}         &\multicolumn{1}{c}{(3)}         &\multicolumn{1}{c}{(4)}         \\
Dep var:            &\shortstack{HH members\\in farm}         &\shortstack{HH hours\\in farm}         & Child labor         &ln(wage bill)         \\
\midrule
Average HDD x Owns livestock&       0.019         &       0.032\sym{*}  &       0.024\sym{*}  &      -0.095         \\
                    &     (0.012)         &     (0.019)         &     (0.012)         &     (0.061)         \\
Average HDD x No livestock&       0.014         &       0.016         &       0.029\sym{*}  &      -0.038         \\
                    &     (0.014)         &     (0.024)         &     (0.015)         &     (0.055)         \\
\midrule
No. obs.            &      26,724         &      26,726         &      14,358         &      53,618         \\
R-squared           &       0.513         &       0.361         &       0.315         &       0.247         \\
\bottomrule
\end{tabular}
}

%% file: content/Table_consumption.tex
{
\def\sym#1{\ifmmode^{#1}\else\(^{#1}\)\fi}
\begin{tabular}{l*{9}{c}}
\toprule
                    &\multicolumn{3}{c}{ln(inc/capita)}                               &\multicolumn{3}{c}{ln(cons/capita)}                              &\multicolumn{3}{c}{Poor (Yes=1)}                                 \\\cmidrule(lr){2-4}\cmidrule(lr){5-7}\cmidrule(lr){8-10}
                    &\multicolumn{1}{c}{(1)}         &\multicolumn{1}{c}{(2)}         &\multicolumn{1}{c}{(3)}         &\multicolumn{1}{c}{(4)}         &\multicolumn{1}{c}{(5)}         &\multicolumn{1}{c}{(6)}         &\multicolumn{1}{c}{(7)}         &\multicolumn{1}{c}{(8)}         &\multicolumn{1}{c}{(9)}         \\
Sample:             &         All         &       Coast         &   Highlands         &         All         &       Coast         &   Highlands         &         All         &       Coast         &   Highlands         \\
\midrule
\hspace{0.1cm} Average DD&       0.023\sym{***}&       0.010         &       0.024\sym{***}&       0.021\sym{***}&       0.013         &       0.021\sym{***}&      -0.014\sym{***}&      -0.009         &      -0.013\sym{***}\\
                    &     (0.004)         &     (0.012)         &     (0.004)         &     (0.004)         &     (0.014)         &     (0.004)         &     (0.003)         &     (0.012)         &     (0.003)         \\
\addlinespace
\hspace{0.1cm} Average HDD&      -0.017         &      -0.015         &      -0.008         &      -0.014         &      -0.016         &       0.001         &       0.003         &       0.009         &      -0.008         \\
                    &     (0.013)         &     (0.013)         &     (0.022)         &     (0.010)         &     (0.010)         &     (0.017)         &     (0.007)         &     (0.008)         &     (0.015)         \\
\midrule
No. obs.            &      53,619         &       7,439         &      46,180         &      53,619         &       7,439         &      46,180         &      53,619         &       7,439         &      46,180         \\
R-squared           &       0.380         &       0.388         &       0.335         &       0.452         &       0.451         &       0.416         &       0.264         &       0.282         &       0.244         \\
\bottomrule
\end{tabular}
}